\documentclass[submitting]{nst}

\usepackage{subfigure,dcolumn}
\usepackage[T2A,T1]{fontenc}
\usepackage[russian,english]{babel}

\usepackage{listings}
\usepackage{multirow}
\begin{document}
\title{Design optimization of the plastic scintillators with WLS-fibers and SiPM readouts in the top veto tracker of the JUNO-TAO experiment}
\thanks{Supported by School of Physics in Sun Yat-sen University, China.}

\author{Guang Luo}
\author{Y.K.Hor}
\email[Y.K. Hor,]{heyuanq@mail.sysu.edu.cn}
\author{Peizhi Lu}
\affiliation{School of Physics, Sun Yat-sen University, Guangzhou 510275, China}
\author{Zhimin Wang}
\email[Zhimin Wang,]{wangzhm@ihep.ac.cn.}
\affiliation{Institute of High Energy Physics, Beijing 100049, China}
\affiliation{University of Chinese Academy of Sciences, Beijing 100049, China}
\author{Ruhui Li}
\affiliation{Institute of High Energy Physics, Beijing 100049, China}
\author{Min Li}
\affiliation{Institute of High Energy Physics, Beijing 100049, China}
\affiliation{University of Chinese Academy of Sciences, Beijing 100049, China}
\author{Yichen Li}
\affiliation{Institute of High Energy Physics, Beijing 100049, China}
\affiliation{University of Chinese Academy of Sciences, Beijing 100049, China}
\author{Liang Zhan}
\affiliation{Institute of High Energy Physics, Beijing 100049, China}
\affiliation{University of Chinese Academy of Sciences, Beijing 100049, China}
\author{Wei Wang}
\email[Wei Wang,]{wangw223@mail.sysu.edu.cn}
\affiliation{School of Physics, Sun Yat-sen University, Guangzhou 510275, China}
\affiliation{Sino-French Institute of Nuclear Engineering and Technology, Sun Yat-sen University, Zhuhai 519082, China}
\author{Yuehuan Wei}
\affiliation{Sino-French Institute of Nuclear Engineering and Technology, Sun Yat-sen University, Zhuhai 519082, China}
\author{Yu Chen}
\affiliation{School of Physics, Sun Yat-sen University, Guangzhou 510275, China}
\author{Xiang Xiao}
\affiliation{School of Physics, Sun Yat-sen University, Guangzhou 510275, China}
\author{Fengpeng An}
\affiliation{School of Physics, Sun Yat-sen University, Guangzhou 510275, China}

\begin{abstract}
Plastic scintillator (PS) embedding wavelength shifting (WLS) fiber is widely used in high energy particle physics, as muon taggers, and also in medical physics and other applications. In this work, a simulation package is built to evaluate the effects of the diameter and the layout of the optical fiber on the light yield with different configurations. The optimal optical configuration was designed based on the simulation and then validated with two PS prototypes under certain experimental conditions. In the study, the top veto tracker (TVT) of the JUNO-TAO experiment, comprised of 4 layers of 160 strips of PS was designed and evaluated. When a muon tagging efficiency of a PS strip is higher than 99\%, the threshold is evaluated. The efficiency of 3-layer out of 4-layer of TVT will be higher than 99\% even with the tagging efficiency of a single strip as low as 97\% using a threshold of 10\,p.e.\,assuming 40\% SiPM PDE.

\end{abstract}


\maketitle

\section{Introduction}
\label{sec:intro}

The collisions between the primary cosmic rays and the earth's atmosphere will produce a large number of muons\cite{muon-rate}, the average kinetic energy of which at the sea level is several GeV\cite{muon-energy}. Because of the high energy, large mass, small deceleration and deflection in the electromagnetic field, and small bremsstrahlung effect with the atomic nuclear electric field in the matter, the muons will have a strong penetration power\cite{Patrignani:2016xqp}. A sub-system of the muon veto detector with high muon tagging efficiency is very important to greatly reduce the background induced by the cosmic-ray (CR) muons for the experiments with only limited overburden near the ground, where the flux of muon is normally four to seven orders of magnitude higher than the underground laboratories with large overburden, such as Jinping underground laboratory\cite{Guo_2021}, Gran Sasso underground laboratory\cite{BARBUTO2004485} and Canfranc underground laboratory\cite{Trzaska2019}. For example, in neutrino experiments\cite{JUNO-detector,JUNO-CDR,TAO-CDR}, dark matter experiments\cite{XENON1T:2014eqx,XENON:2020kmp,MAGIX:2022fdt,DarkSide:2013syn}, neutrino-less double beta decay experiments\cite{Pocar:2015ota,Tosi:2013bta,Gornea:2009zz}, those muon veto systems require muon tagging efficiency higher than 99$\%$. At present, the detectors based on plastic scintillator (PS) have the advantage of easy machining\cite{Birks:1964zz,Zhezher:2020qov,NUCLEUS:2022vyj,Seo:2022vzr,veto-LBNL-THOMAS201347}, flexible structure design, efficient and stable performance\cite{Pla-Dalmau:2000puk,Moiseev:2007zz}. The PS detectors\cite{Vaishali2021Design}, especially with WLS fibers\cite{Holm:1989sb,Bloise:2023xhc,Buzhan:2020ryp,Bugg:2013ica,Jia-Ning2018Position-sensitive} and optical photodetectors (Multi-anode PMTs or Silicon PhotoMultipliers(SiPMs)), were used in OPERA\cite{Adam:2007ex}, MINOS\cite{MINOS:2002xlc}, LHAASO\cite{Wang:2021ejh,LHAASO:2021awk} and many other experiments\cite{Evans:2013pka,Orsi:2007zz,Andreev:2004uy,Thompson:2022ufx}. Meanwhile, the PS detectors have many applications in geological imaging\cite{PROCUREUR2018169,Morishima:2017ghw,Zenoni:2014kva,Marteau:2012zv}, reactor monitoring and other fields\cite{Oguri:2014gta,Georgadze:2016ufb,Scovell:2013hva} . 

The Taishan Antineutrino Observatory (TAO or JUNO-TAO) is a satellite experiment of the Jiangmen Underground Neutrino Observatory (JUNO)\cite{TAO-CDR,JUNO-CDR}. The main purpose of the TAO experiment is to provide a precise neutrino energy reference spectrum for JUNO and benchmark measurements for the nuclear database. TAO detector system will consists of a central detector (CD), an outer shielding and veto system. The CD will be placed at around 30\,m from one core of the Taishan Nuclear Power Plant. The CD consists of a 2.8-ton gadolinium-doped liquid scintillator (LS) filled in a spherical acrylic vessel. The gadolinium-doped LS, as a target material, reacts with neutrinos from the reactor to measure the neutrino energy spectrum\cite{Capozzi:2020cxm}. Since TAO only has a limited overburden with 4 meter, the major backgrounds for the TAO experiment are muon spallation products and accidental coincidences, mostly due to the natural radioactivity, 
the top veto tracker (TVT) require to tag muons with efficiency higher than 99\%.

In this paper, a comparison is realized in sec.\ref{sec:Exp} between a simulation based on Geant4\cite{geant4-AGOSTINELLI2003250,Riggi:2010zz,Wenzhen2013Geant4} and a measurement of a prototype of PS strip with WLS-fiber readout. The light yield results of the experiment and simulation are consistent for passing through muons. In sec.\ref{sec:Opt}, the diameter and the layout of the WLS fiber were further checked for higher light yield against with the simulation. An optimized design of the PS strip with WLS-fiber and SiPM readout is proposed for the TVT system of JUNO-TAO with high light yield and muon tagging efficiency, which will provide a good reference and guidance for the design of PS detector with WLS-fiber. At the same time, the reliability of the optimal design is preliminarily proved by experiments. In sec.\ref{sec:TPS}, with the proposed PS strip design, the expected performance of the TAO TVT system is demonstrated. Finally, a summary is given in sec.\ref{sec:Sum}.

\section{Prototype of PS strip with WLS-fiber and simulation}
\label{sec:Exp}

Muons will deposit their energy when they pass through and interact with the surrounding materials, and the process of muon energy loss is called muon ionization energy loss\cite{Patrignani:2016xqp}. The average energy loss per distance (mass thickness) can be described by the Bethe-Bloth formula (\ref{equation:one}) \cite{Patrignani:2016xqp,Lecoq:2020itu}:

{\small
\begin{eqnarray}
 \label{equation:one}
 -\frac{dE}{dx}=Kz^{2}\frac{Z}{A}\frac{1}{\beta^{2}}[\frac{1}{2}ln(\frac{2m_{e}c^{2}\beta^{2}\gamma^{2}W_{max}}{I^{2}})-\beta^{2}-\frac{\delta}{2}],
\end{eqnarray}
}

where $K$ is a constant, $z$ is the unit charge of the incident muon, $m_{e}$ and $c$ are the electron mass and the speed of light, respectively. $Z$ and $A$ are the atomic number and mass number of the passing-through matter. $W_{Max}$ is the maximum kinetic energy that can be transferred to an electron when the muon collides with the atom. $I$ is the average excitation energy of the matter. $\beta$ is the ratio of the speed of particle to the speed of light. $\gamma$ is lorentz factor. $\delta$ is the correction factor of the density effect of the matter. The above parameters are constant for a given matter.

From the formula (\ref{equation:one}), the deposited energy of muon in the material is related to the energy of muon and atomic number of the material. For thin-layer media with an atomic number less than 20, such as a PS strip, muons almost pass through in a straight line. Partially lost energy of muon will be converted into light in a PS strip, and the light will be exported by the WLS-fiber, where SiPMs coupled with the fiber is an effective, convenient, and rapid method to pick up the photons and then convert them into electrical signal. 

\begin{figure}[!htbp]
\begin{center}
\includegraphics[scale=0.4]{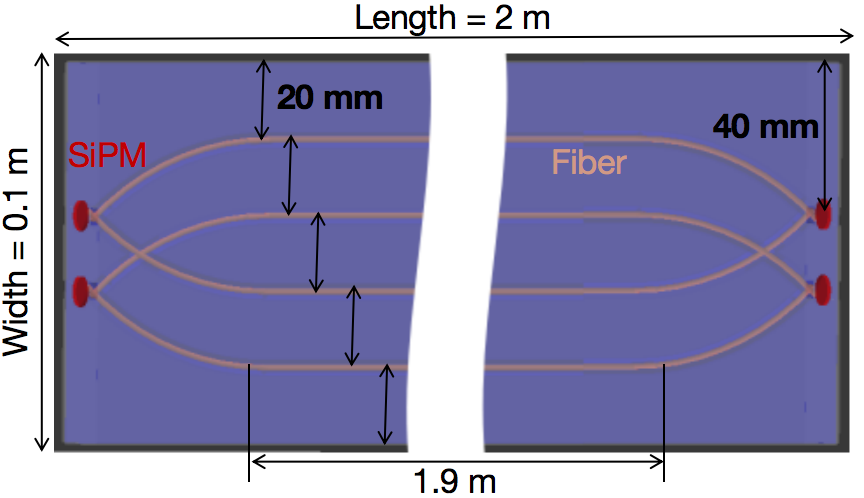}
\caption{Design of a prototype of the PS strip with WLS-fiber (Subsequently, it is named as option 1)}
\label{Prot}
\end{center}
\end{figure}

A prototype of the PS with WLS-fiber readout is designed and fabricated as shown in Figure \ref{Prot}\cite{Min2023Performance}, named option 1. Its size is in a dimension of 2\,m (Length)$\times$0.1\,m (Width)$\times$0.02\,m (Thickness), and four optical fibers with a diameter of 1\,mm are used in total. The pink lines represent the WLS fiber equally spaced, inserted, and filled in the surface of the PS strip. The length of the straight length part is 1.9\,m. The arrangement is symmetrical in both length and width directions. Two of the fibers are focused into a single group which can be coupled with optical sensors. This option can reduce the number of optical sensors. For example, four SiPMs (red circular point in the figure) can be used for each fiber group, or two PMTs can be used for each PS end. Finally, the PS is wrapped in reflective film (aluminum foil) except for the pips of the optical fiber to export photons, and more details can be found in Ref\cite{Min2023Performance}. The current prototype was tested with CR muon before design optimization.

\begin{figure*}[!htbp]
\begin{center}
\includegraphics[scale=0.4]{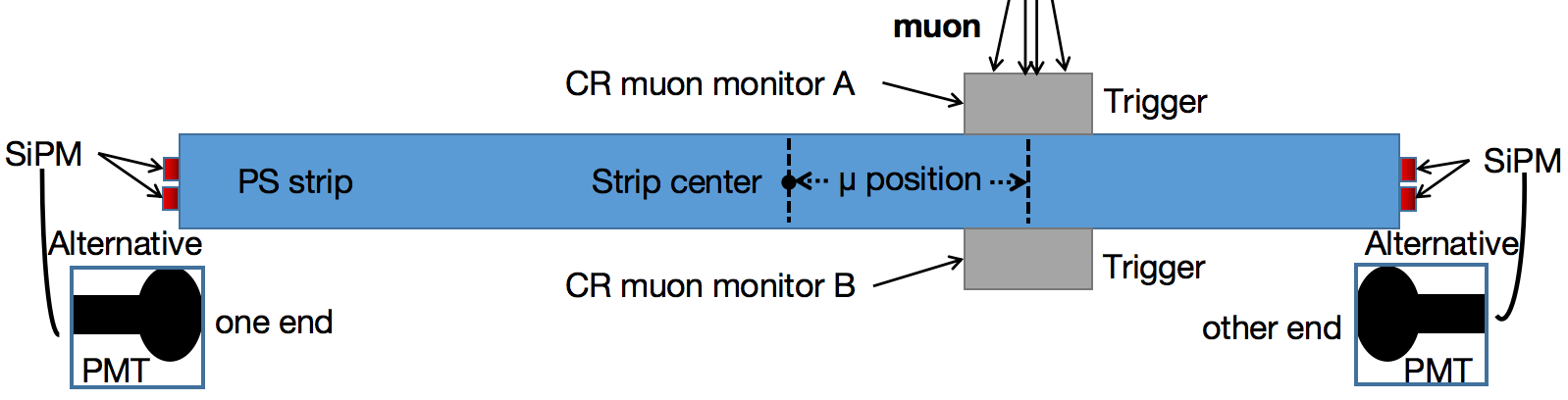}
\caption{Device and principle diagram for CR muon measurement with the PS strip prototype}
\label{EX}
\end{center}
\end{figure*}

A CR muon survey along the length of PS strip prototype with PMT/SiPM is done with the schema shown in Figure \ref{EX} (the arrangement of optical fibers is not shown). The CR muon\cite{Bugg:2013ica,Yang:2022awv} will be selected by the two muon monitors of 
small scintillators on different hitting locations. When the CR muon monitors are triggered, the signals of the PS strip will be recorded too. 9 points equally spaced along the strip were measured in total relative to the center of the PS strip. More details about the experiment can be found in Ref \cite{Min2023Performance}. The PS strip is fabricated by Beijing Hoton Nuclear Technology Co., Ltd \cite{Haotang:2014}. The type of fiber is the WLS fiber BCF92 \cite{Tur:2009en,Dietz-Laursonn:2016tpy}. 
For further understanding and optimization, a Monte-Carlo (MC) simulation project is set up based on Geant4\cite{geant4-AGOSTINELLI2003250,Yang:2022awv,Wenzhen2013Geant4} with the parameters provided by the manufacturer of the PS, the reflective film and WLS-fiber \cite{Dietz-Laursonn:2016tpy,Qian:2021jlv}. 
The simulation project mainly includes three parts: The first part is the detector geometry, the PS geometry with optical fibers are  designed through a geometric interface. This is why there are various geometric designs in subsequent optimizations. The second is the physical process section, which contains a physical list and the optical processes of optical photons. The physical list includes such things as ionization, bremsstrahlung, multiple scattering, pair generation, Compton scattering, and photoelectric effects. Optical processes include the generation of Scintillation and Cherenkov light, wavelength shift effects, Rayleigh scattering, bulk absorption, and boundary processes. The third is the extraction and analysis of information. The PS and SiPM are set as sensitive areas; The PS is responsible for obtaining information about the muon, and The SiPM is responsible for obtaining information about the photons hitting it. A parameter interface is provided in the simulation package to set the properties of the material, such as PS attenuation length, scintillator yield (refers to the number of photons converted when the energy deposited in the scintillator is 1 MeV), reflectivity of reflective film. By scanning these parameters, we obtain a series of simulation responses i.e. light yield (refers to photoelectron (p.e.) with the consideration of the corresponding photon detection efficiency(PED)/quantum efficiency(QE) of SiPM/PMT ) along PS longitudinal direction, and perform $\chi^2$ analysis with experimental data. Table \ref{table1} is a list of parameters corresponding to the minimum $\chi^2$.
In the subsequent optimization work, The parameters of PS, optical fiber and reflective film in table \ref{table1} are the same in simulation. 
\begin{table}[!htbp]
 \centering
 \caption{The key parameres of the simulation of the PS strip prototype}
 \label{table1}
 \begin{tabular}{|c|c|c|}
 \hline
 Material & Properties & Parameters  \\ \hline
\multirow{4}{*}{PS} & Base & Polyvinyltoluene   \\ \cline{2-3}
\multirow{4}{*}{} & Scintillation yield (photons/MeV) & 8000 \\ \cline{2-3}
\multirow{4}{*}{} & Emission peak (nm) & 415 \\ \cline{2-3}
\multirow{4}{*}{} & Attenuation length (cm) & 200 @ 400 nm \\ 
\hline
\multirow{2}{*}{WLS-Fiber} & Core & Polystyrene  \\ \cline{2-3}
\multirow{2}{*}{ } & Attenuation length (cm) & 380 @ 400 nm \\
  \hline
\multirow{2}{*}{Reflective film} & Base & Aluminium  \\ \cline{2-3}
\multirow{2}{*}{ } & Reflectivity & 85\% \\
   \hline
 \end{tabular}
\end{table}

\begin{figure*}
   \subfigure[]{
   \begin{minipage}[t]{0.45\linewidth}
   \centering
   \includegraphics[width=9cm]{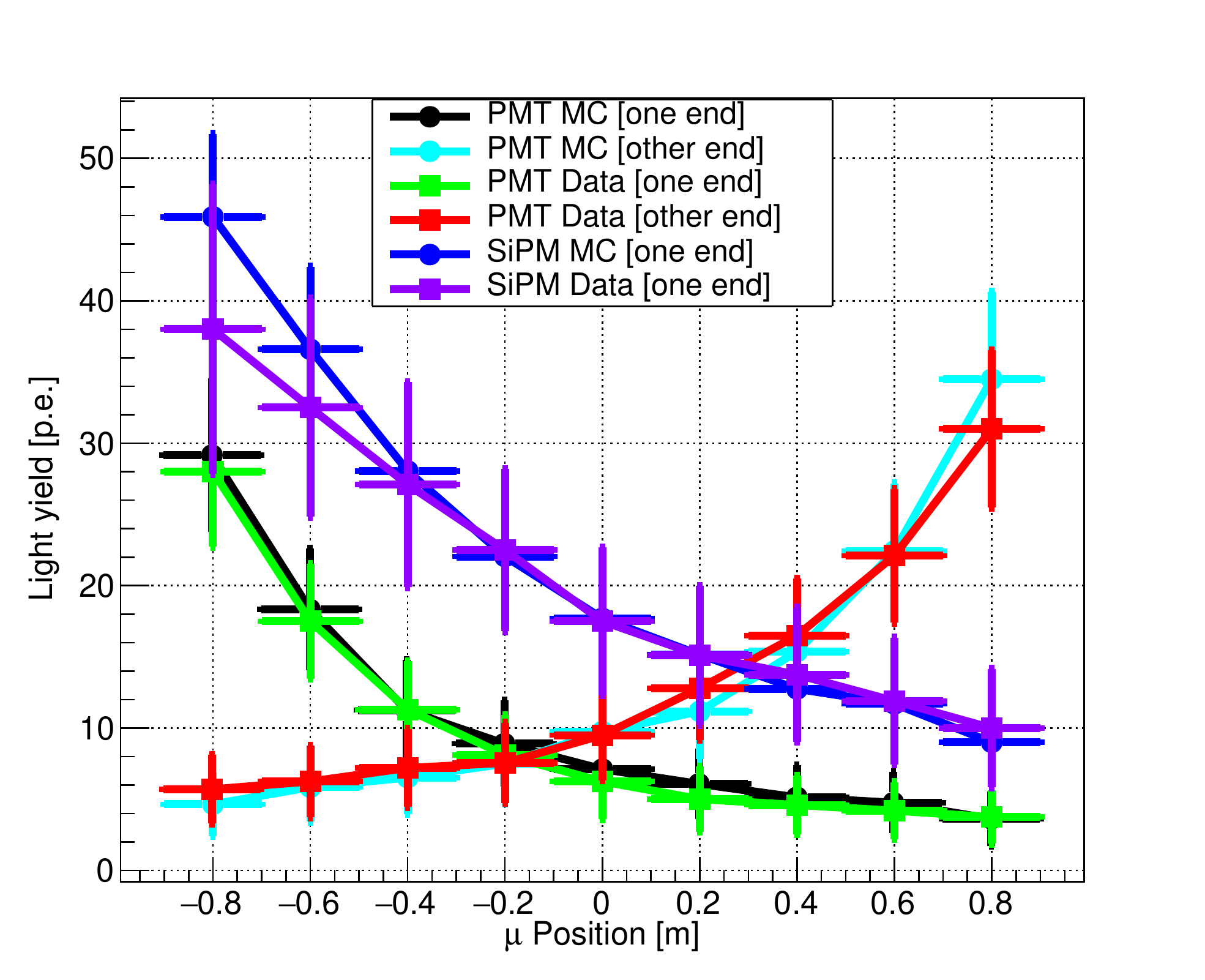}
    \label{a}
    \end{minipage}
    }
    \subfigure[]{
    \begin{minipage}[t]{0.45\linewidth}
    \centering
    \includegraphics[width=9cm]{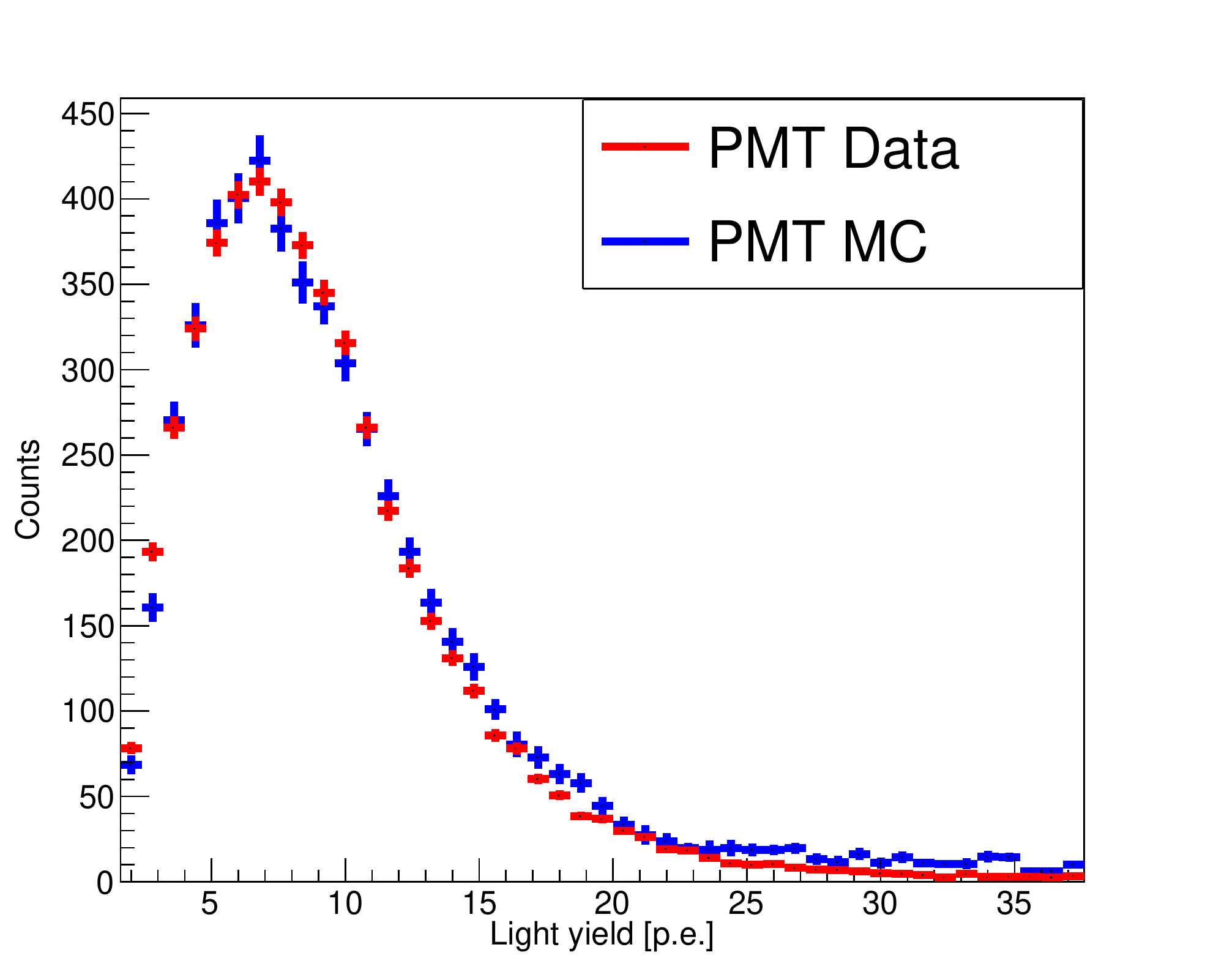}
    \label{b}
    \end{minipage}
    }
    \caption{(a):Comparison of light yield between measurement and simulation at different muon hitting positions. The horizontal error bar is from the dimension of the muon monitors, the vertical error bar in MC is the statistical error. Due to the different QE of PMT, light yield at the two ends are not strictly symmetrical. (b) When the CR muon monitor in the center of PS, Comparison diagram of experimental energy spectrum and simulation of PMT at one end.}
    \label{compari}
\end{figure*}

The prototype measurements of the survey have been simulated. The comparison of light yield between the measurement and simulation at different positions is shown in Figure \ref{a}, where the light yield (in p.e.) refers to the average value of the photoelectron distribution of the selected muons hitting each position and the X-axis represents the distance from the center of the PS strip in the length direction. When using PMT as sensor, the simulation and data are in good agreement within the error range. The light yield at both ends should be symmetrically distributed around the center of the PS. Due to the different QE of PMT, light yield at the two ends are not strictly symmetrical. When using SiPM of the same type as a sensor,The distribution of light yield at one end is shown here. Within the range of error, the simulation and experiment in most areas of PS are consistent, while the compliance is poor in the areas at the edge of PS. As can be seen from the figure, the light yield of SiPM is larger than that of PMT as a sensor.
In addition to the initial comparison shown in Figure \ref{a}, Similarly,A comparison of energy spectra is also obtained, Comparison diagram of energy spectra between the measurement and simulation at center positions of PS is shown in Figure \ref{b},Except for individual energy points that do not conform to the experiment, others are in good agreement with the experiment within the error range.
From an experimental perspective, each PMT/SiPM has a different QE/PED, Since the following optimization work does not pay attention to the impact of electronics, In subsequent studies, the QE of each PMT or the PDE of SiPM will be input with the same value.
\section{Optimization of PS strip layout}
\label{sec:Opt}
\begin{figure*}[!htbp]
\begin{center}
\includegraphics[scale=0.4]{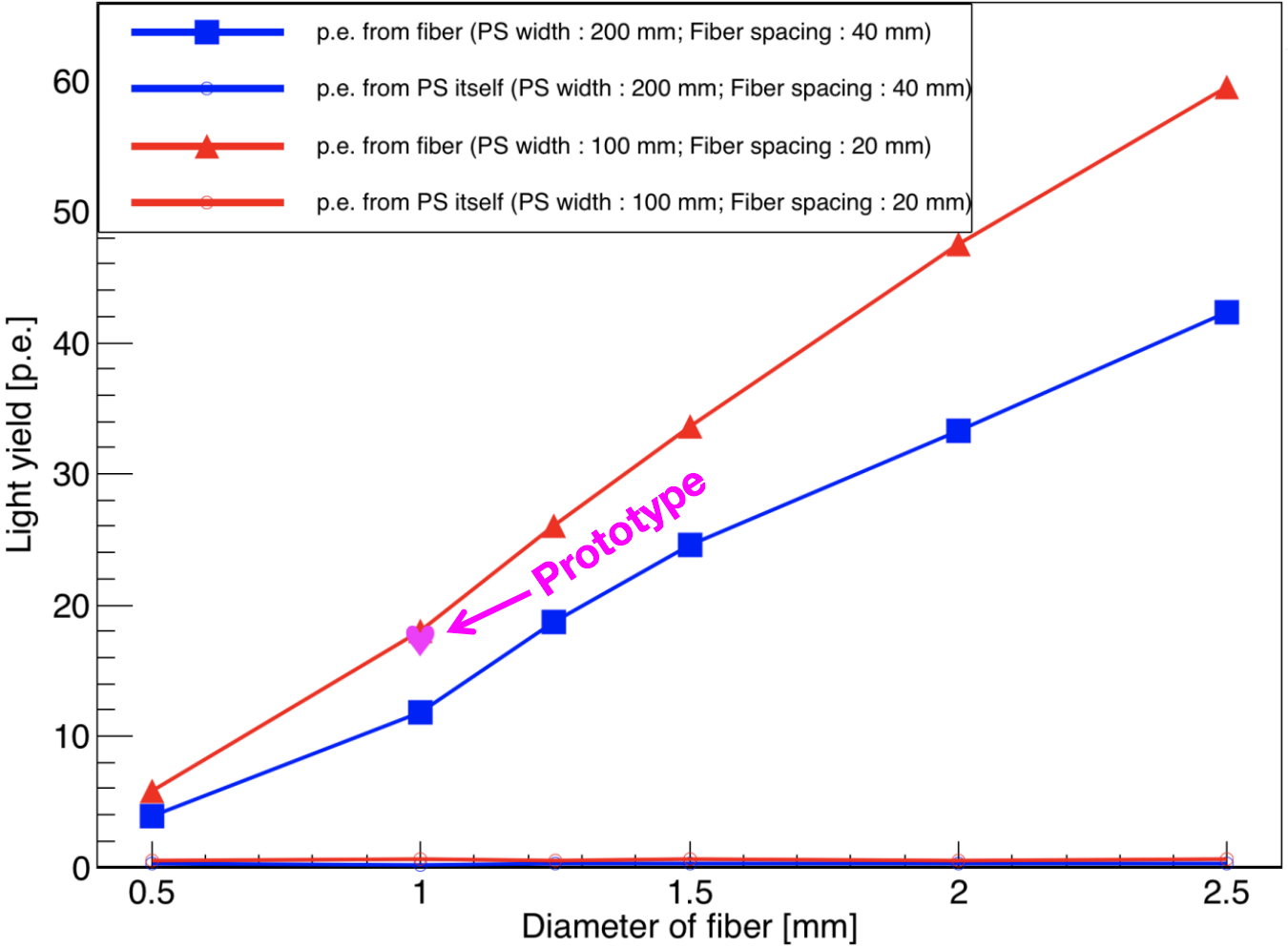}
\caption{Assuming 30\% SiPM PDE; Each point refers to the light yield of the muon hitting the PS center.} Two widths of 200 \,mm and 100 \,mm for the PS are checked with the same length of 2 \,m and thickness of 0.02 \,m. The same four fibers are used in each case, which is why the light yield of the 200 \,mm wide PS is smaller than that of the 100 \,mm one.
\label{GsZ}
\end{center}
\end{figure*}
\begin{figure*}[!htbp]
\begin{center}
\includegraphics[scale=0.4]{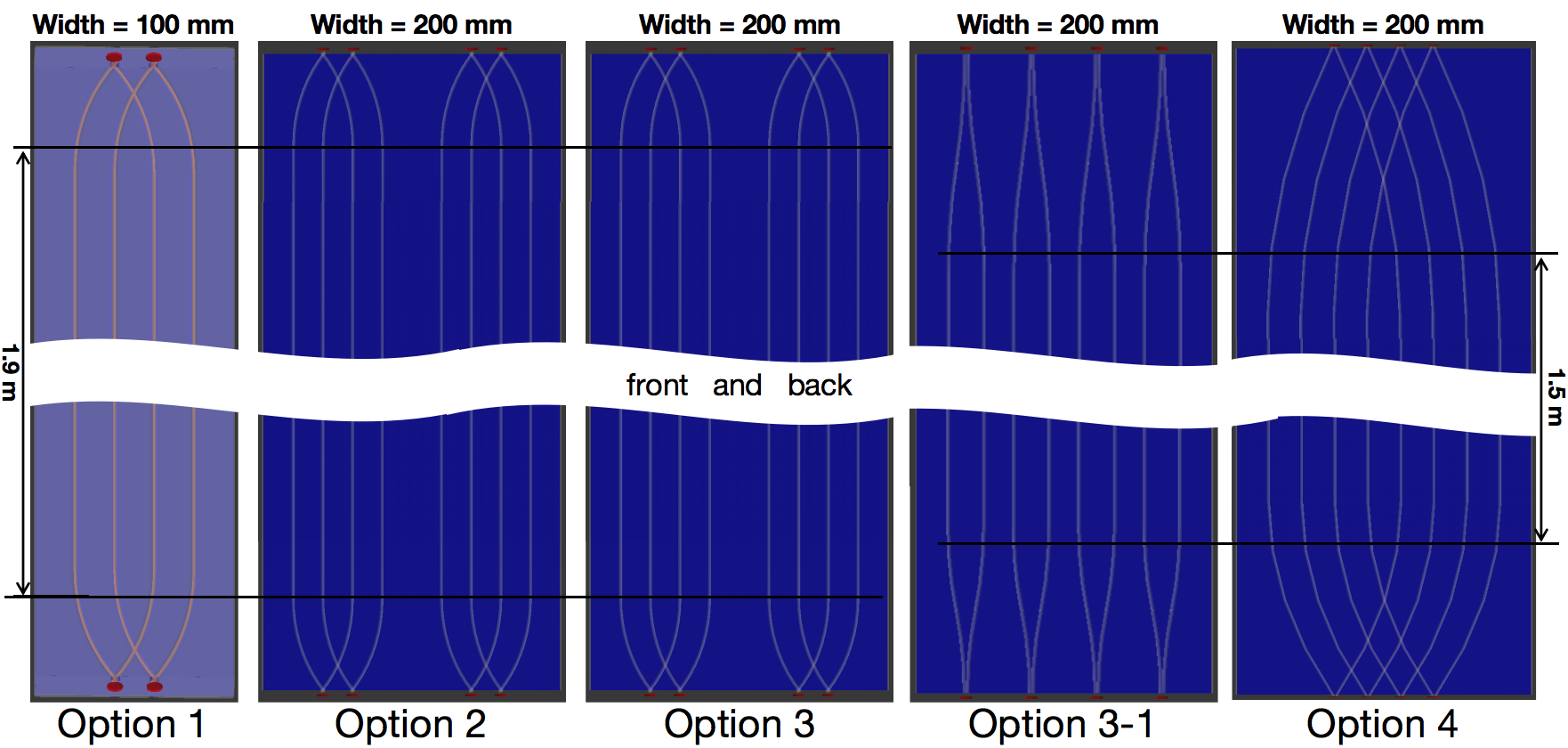}
\caption{All evaluated designs of a single PS strip.The number of fibers is eight to have a similar spacing except option 1. The fiber of option 2 is only arranged on the front of the PS as doubled of option 1. Option 3 is similar to option 2, but with four optical fibers installed on the back and another four on the front of the PS, respectively.} 
\label{figSPS}
\end{center}
\end{figure*}
 On the basis of the agreement between experiment and simulation, a further study on the configurations of the PS strip design is done including the diameter and the arrangement of the WLS-fiber for the light yield and tagging efficiency.
The relationship between the light yield and fiber diameter is shown in Figure \ref{GsZ} for 2 PS configurations with the same number of fibers embedded. Each point refers to the light yield of the muon hitting the PS center. The magenta heart represents the measurement of the prototype mentioned earlier. The red lines represent the simulated relationship between the light yield and the fiber diameter with the PS width of 100\,mm, where the spacing between neighbor fibers is around 20\,mm. The blue is the simulated relationship between the light yield and the fiber diameter with the PS width of 200\,mm, where the spacing between neighbor fibers is around 40\,mm. Most photons are collected through the WLS fiber by the sensors. Since the area of the optical sensor (Since TAO TVT requirement for readout, subsequent optical sensors default to SiPMs) is larger than the dimension of the optical fiber, there are still some PS scintillation photons directly collected by the sensors without going through the WLS fiber. The two straight lines at the bottom represent the photons directly from the PS scintillation to the SiPM, where the light yield contributed by the PS itself is basically independent of the fiber diameter. 

According to the trend of the plot, the larger the fiber diameter, or the smaller spacing of the neighbor fibers, the higher the light yield. Finally, a 1.5\,mm fiber diameter is suggested according to the expected response and reasonable cost. 

Following the requirements of the JUNO-TAO TVT system, the PS strip with 20\,cm width is suggested according to the fabrication, electronics, and cost, but there is still more than one option proposed following a different strategy as shown in Figure \ref{figSPS}. 
The length and thickness of all the options are the same as 2\,m, and 20\,mm as a basic requirement. The width of option 2, option 3, option 3-1, and option 4 are 200\,mm, while the width of option 1 is only 100\,mm as the measured prototype. The number of fibers is eight to have a similar spacing 
except option 1. The fiber of option 2 is only arranged on the front of the PS as doubled of option 1. Option 3 is similar to option 2, but with four optical fibers installed on the back and another four on the front of the PS, respectively. The fiber layout of option 3-1 and option 4 are in a different arrangement. The fiber of options 3-1 is more uniform than that of option 2, but it is not staggered like that of option 4. For option 1, option 2, and option 3, the length of the straight portion of the fiber is 1.9\,m. For option 3-1 and option 4, the length of the straight portion of the fiber is 1.5\,m.

\begin{table}[!htbp]
 \centering
 \caption{Main configurations of different options of PS strip layout}
 \label{table2}
 \resizebox{0.48\textwidth}{10mm}{
 \begin{tabular}{|c|c|c|c|c|c|}
 \hline
  Configuration &  option 1 & option 2 &option 3 & option 3-1 &option 4  \\\hline
  PS width (mm) & 100 & 200 & 200 & 200 & 200 \\\hline
  Fiber numbers & 4 & 8 & 8 & 8 & 8 \\\hline
  Fiber diameter (mm) & 1 & 1.5 & 1.5 & 1.5 & 1.5\\\hline
  Fiber spacing (mm) & 20 & 20 & 20 & 24 & 22.5 \\\hline
\end{tabular} }
\end{table}

The PS width and WLS fiber diameter of all the options are listed in table \ref{table2}.
The performance (mainly on the light yield and muon tagging efficiency) and differences among the proposals were further evaluated by simulation. It can be divided into three categories for comparative study.
\begin{description}
\item[For Option 2 VS.\,Option 3]
 {addresses the dependence of light yield on fiber placement.}
\item[For Option 2 VS.\,Option 3-1]
 {addresses the basic dependence 
of light yield on the uniformity of the fiber arrangement 
in the PS}
\item[For Option 2 VS.\,Option 4]
 {addresses the dependence of light 
yield on the fiber layout.}
\end{description}

\subsection{Light yield}
\begin{figure*}
   \subfigure[]{
   \begin{minipage}[t]{0.42\linewidth}
   \centering
   \includegraphics[width=8.5cm]{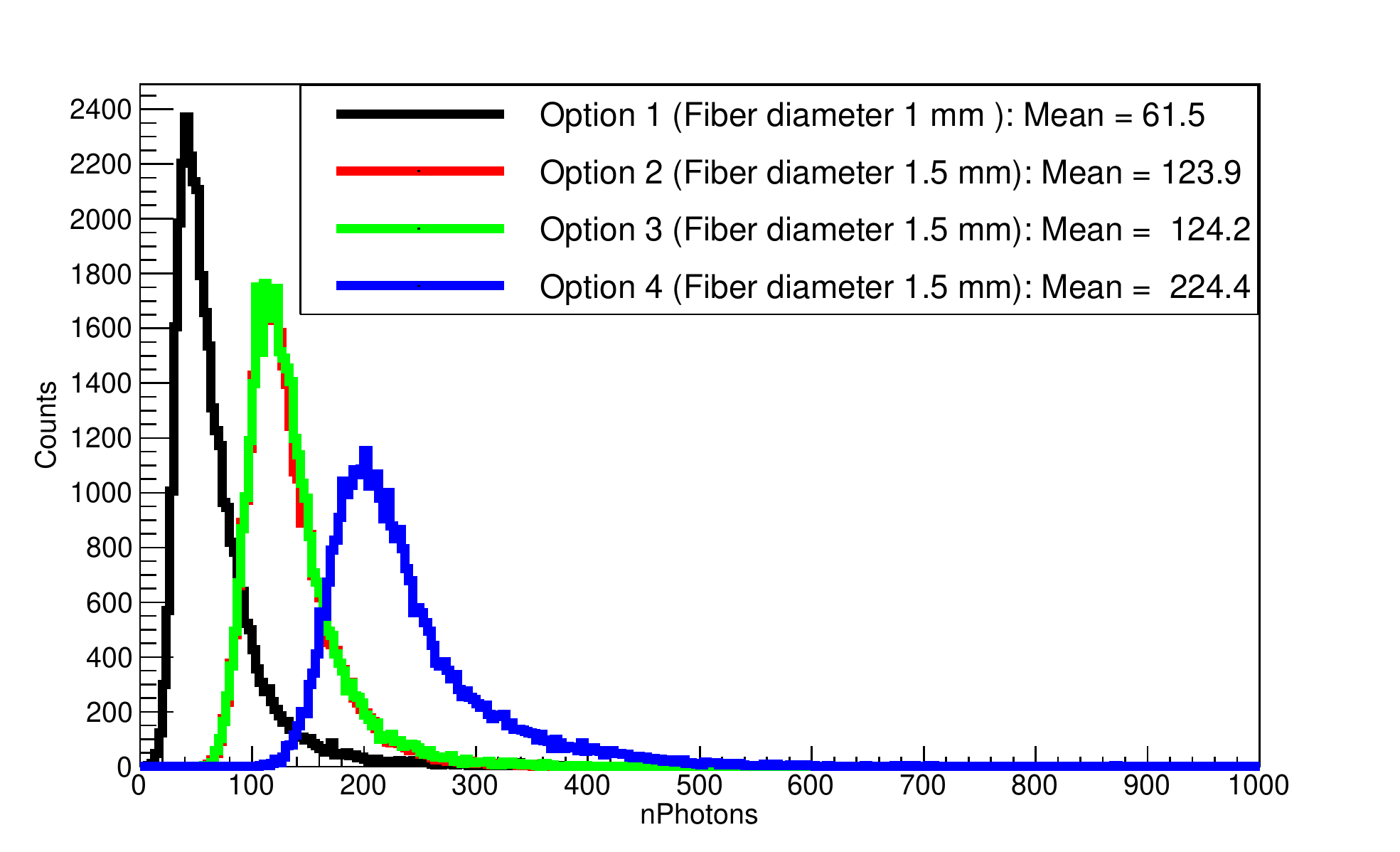}
    \label{4a}
    \end{minipage}
    }
    \subfigure[]{
    \begin{minipage}[t]{0.45\linewidth}
    \centering
    \includegraphics[width=9cm]{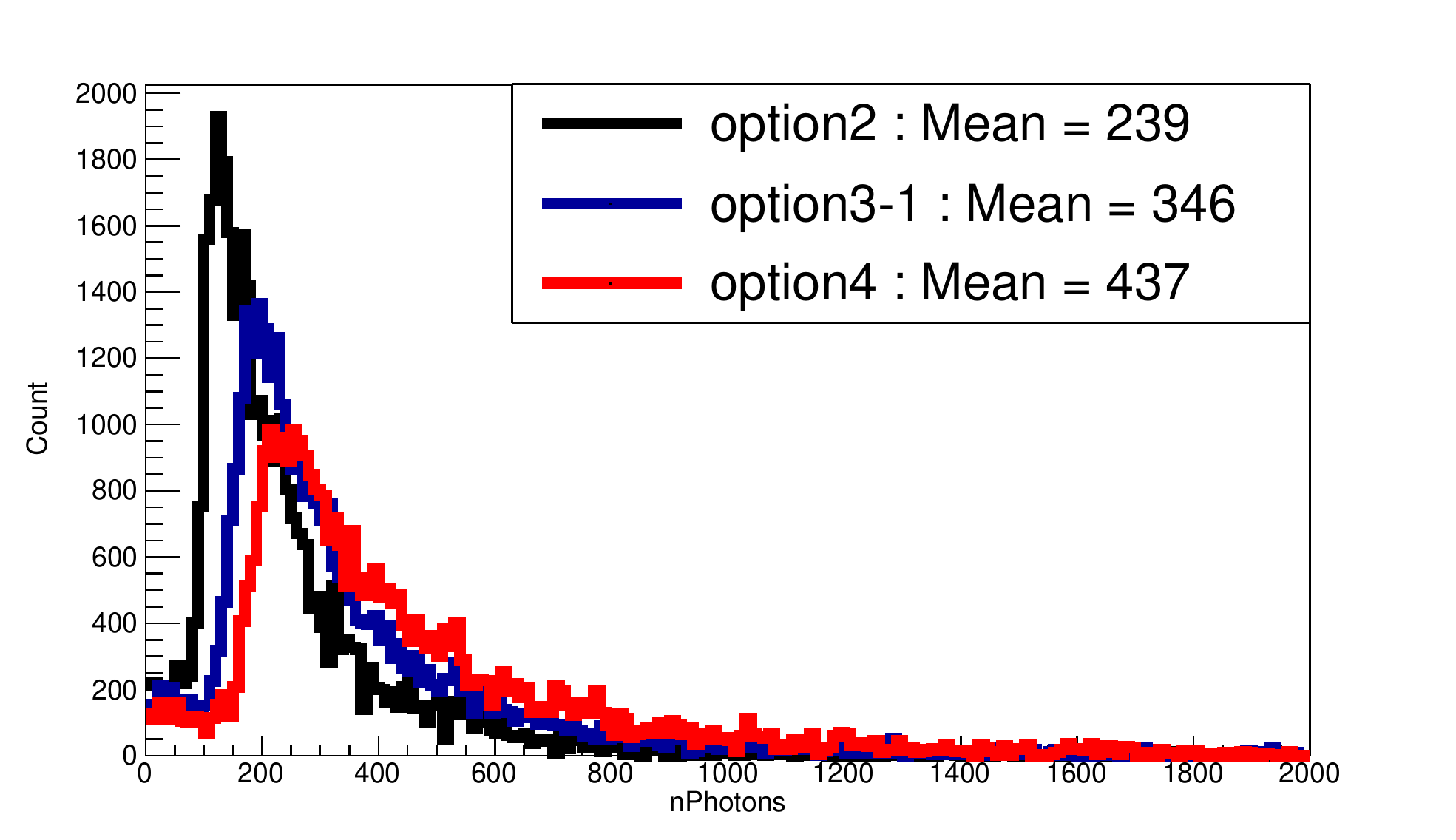}
    \label{4b}
    \end{minipage}
    }
    \caption{(a):With vertical incident muon simulation: Distribution of total number of photons received by SiPM for four different options of PS strip layout. nPhotons:the number of photons received/collected by the SiPM, without the consideration of the corresponding PDE of SiPM(b):With realistic muon simulation: Distribution of total number of photons received by SiPM with different fiber arrangement. nPhotons:the number of photons received/collected by the SiPM, without the consideration of the corresponding PDE of SiPM}
    \label{MuonSim}
\end{figure*}
The angular distribution of the CR muon hitting the PS will affect the PS response. To check the difference and eliminate the additional influence of angular dependence, we use the muons that are vertically and uniformly incident on the PS strip in the simulation firstly. 
The results of the number of photons received by SiPM for option 1-4 are shown in Figure \ref{4a}. The X-axis is the number of photons (nPhotons:the number of photons received/collected by the PMT/SiPM, without the consideration of the corresponding PDE/QE of SiPM/PMT), light yield = nPhotons $\times$ (PED/QE) : the sum of the photons collected by all the SiPMs in a muon event. The figure also shows the average number of photons received by all SiPMs for all muon events. The black line is the distribution of the photons collected by SiPM of option 1. The red and the green lines represent the photon number distributions of option 2 and option 3, which overlap, indicating that the back or front location of fibers has no obvious effect on the collection of the photons generated by the muons. However, the specific reasons are not clear, including the absorption and reemission of PS, or wavelength shift effects, reflection, and other reasons that can lead to loss of directionality. The number of photons in option 2 is almost twice that of option 1. Option 4 has more photons output than the previous three options, which indicates that the arrangement of fibers has a large impact on the light yield. The difference of photon output between option 4 and option 2 is studied in detail in section \ref{subsec:TE}.  All the plots have the same entries, therefore the maximum height of the plots are related to the distribution width.

\begin{figure*}[!htbp]
\begin{center}
\includegraphics[scale=0.42]{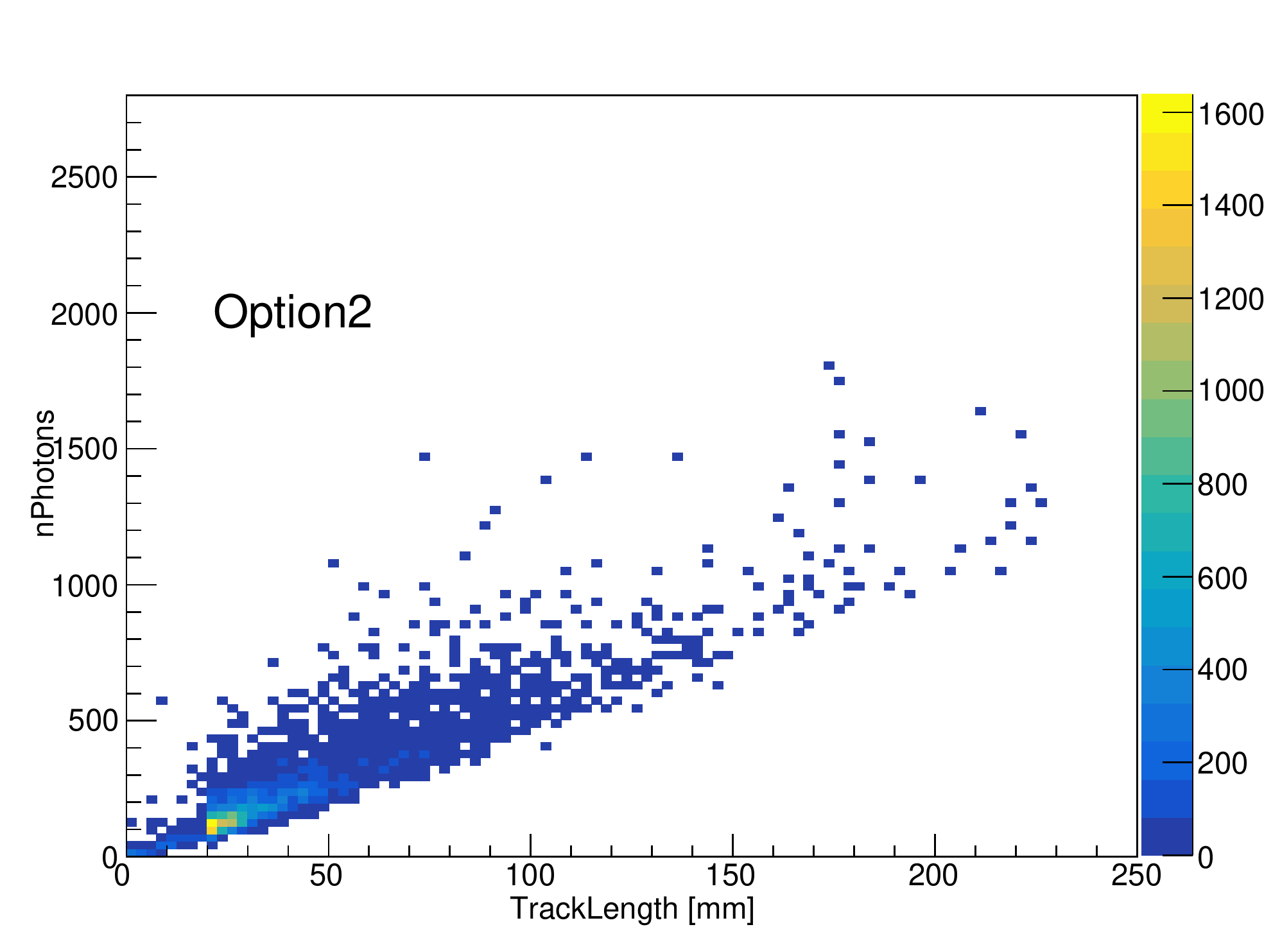}
\includegraphics[scale=0.42]{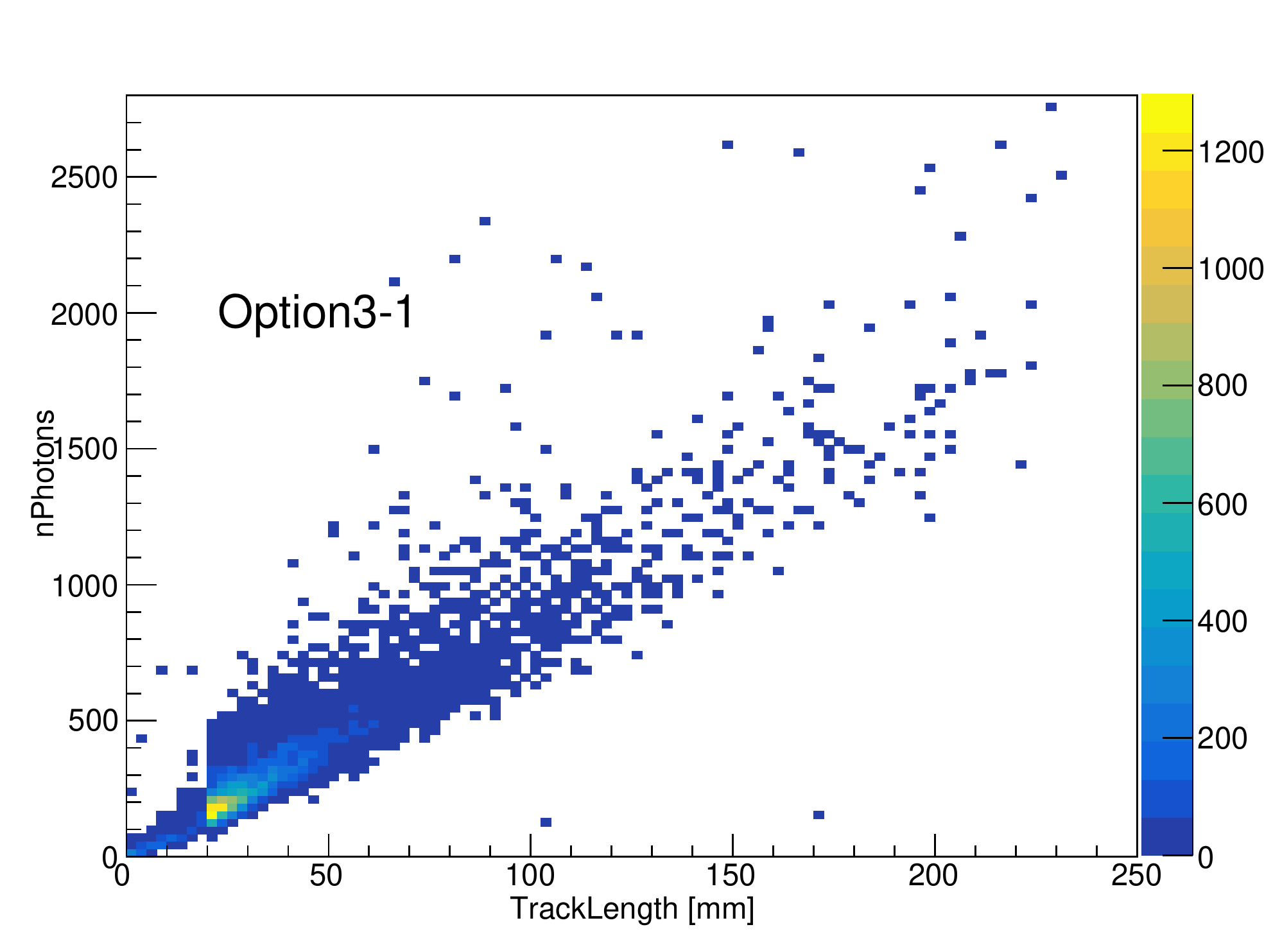}
\includegraphics[scale=0.42]{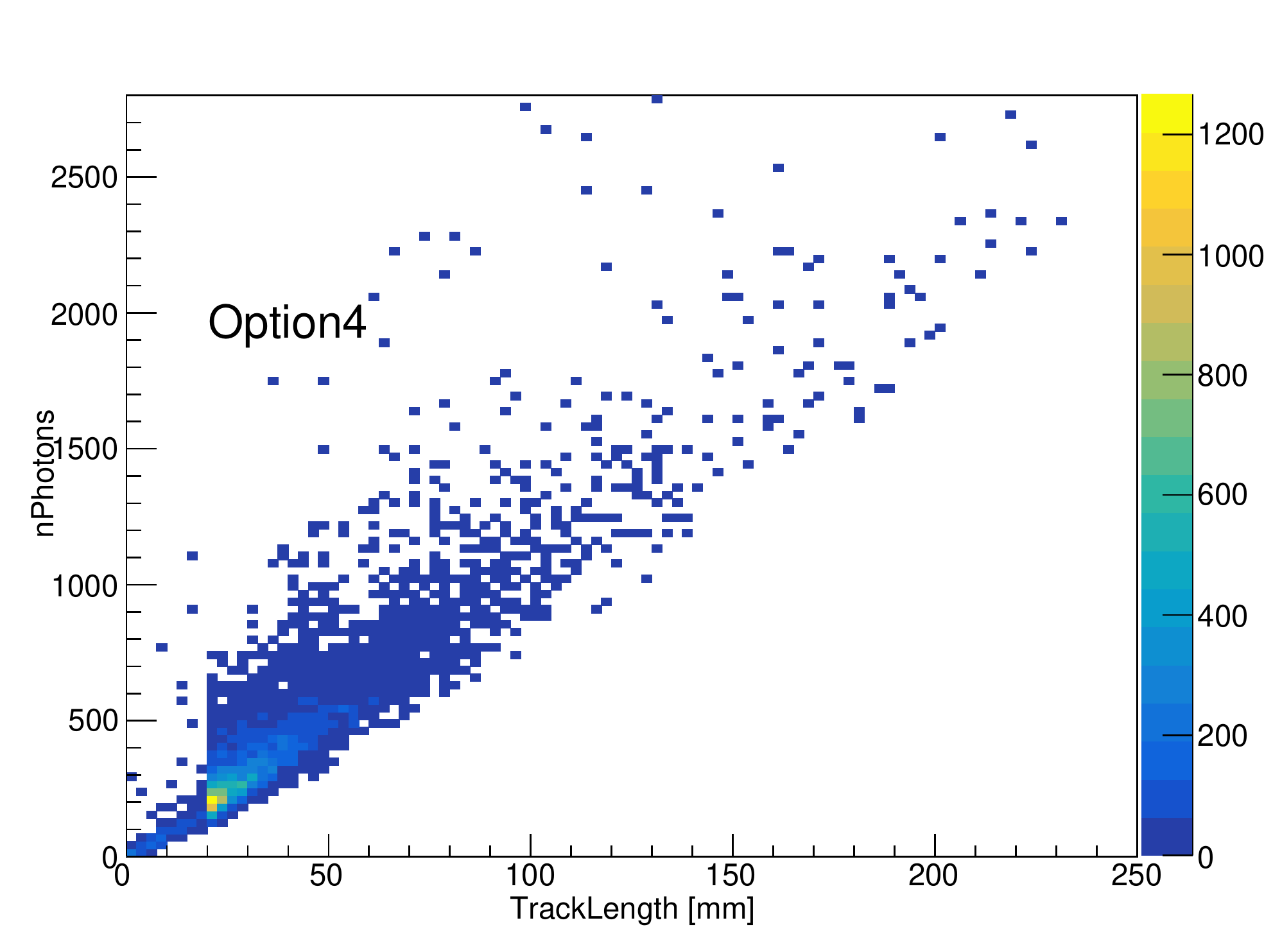}
\includegraphics[scale=0.44]{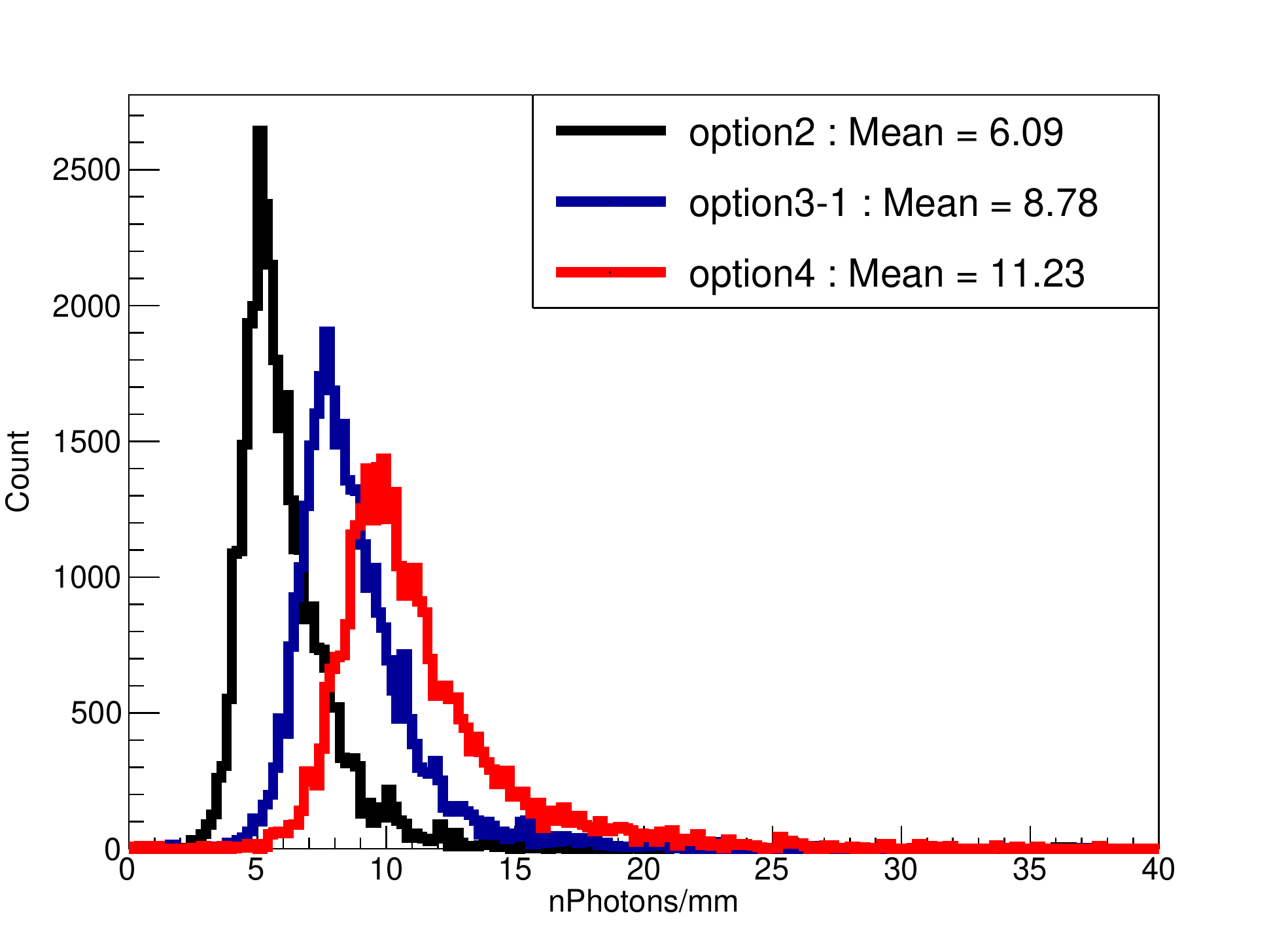}
\caption{Total photon number versus muon track length. Upper left, upper right, bottom right of the picture represent the two-dimensional diagram of total photon number versus muon track length for option 2, option 3-1, option 4 respectively. Bottom right: comparison of the photon number normalized to unit muon track length. nPhotons:the number of photons received/collected by the SiPM, without the consideration of the corresponding PDE of SiPM}
\label{2D234}
\end{center}
\end{figure*}

Another simulation is done to evaluate the differences between option 2, option 3-1, and option 4. The muon generator here is updated in energy and angular distribution according to the Ref.\cite{Gaisser:2016ddr,Gaisser:2016uoy,Shukla:2016nio} as a more realistic situation to model the response of the PS strip. The results are shown in Figure \ref{4b}. The black, blue, and red lines represent the total photon number distribution of option 2, option 3-1, and option 4, respectively. The figure also shows the average value of photons received by all SiPM of all muon events. Except for the total number of photons, the average value of photons of option 4 is the highest, option 3-1 is the middle, and option 2 is the lowest. Option 4 has the best ability to collect photons.

From the formula (\ref{equation:one}), we can know that the deposited energy is directly proportional to the track length of muon passing through the material.
Because of the oblique incidence of muons of the updated muon generator in the second simulation, the track length of the muons in the PS strip can be either less(represent an edge events) or greater than the thickness of the PS strip. The total photon number in Figure \ref{4b} is much higher than that in Figure \ref{4a} even with the same option: the average value of photons has nearly doubled, while some signals with small amplitude show up.
The upper left panel of Figure \ref{2D234} shows the two-dimensional diagram of total photon number versus muon track length for option 2, the upper right panel is for option 3-1, and the bottom left panel is for option 4. The colors of the three panels represent the density of events. The total photon number is proportional to the track length as expected. When the track length is 20\,mm (thickness of the PS), the event density is the highest due to the maximum flux density when muon is vertically incident. The bottom right panel of Figure \ref{2D234} shows the photon number per millimeter of muons passing through the PS strip of the three options. It is obvious that the average value of photons per unit length (mm) of option 4 is 11.23, which is nearly twice that of option 2. This concludes that option 4 is the most effective one for the light yield response to muon when the muon deposits the same amount of energy among these options.

\subsection{Transmission performance}
\label{subsec:TE}
\begin{figure}[!htbp]
\begin{center}
\includegraphics[scale=0.35]{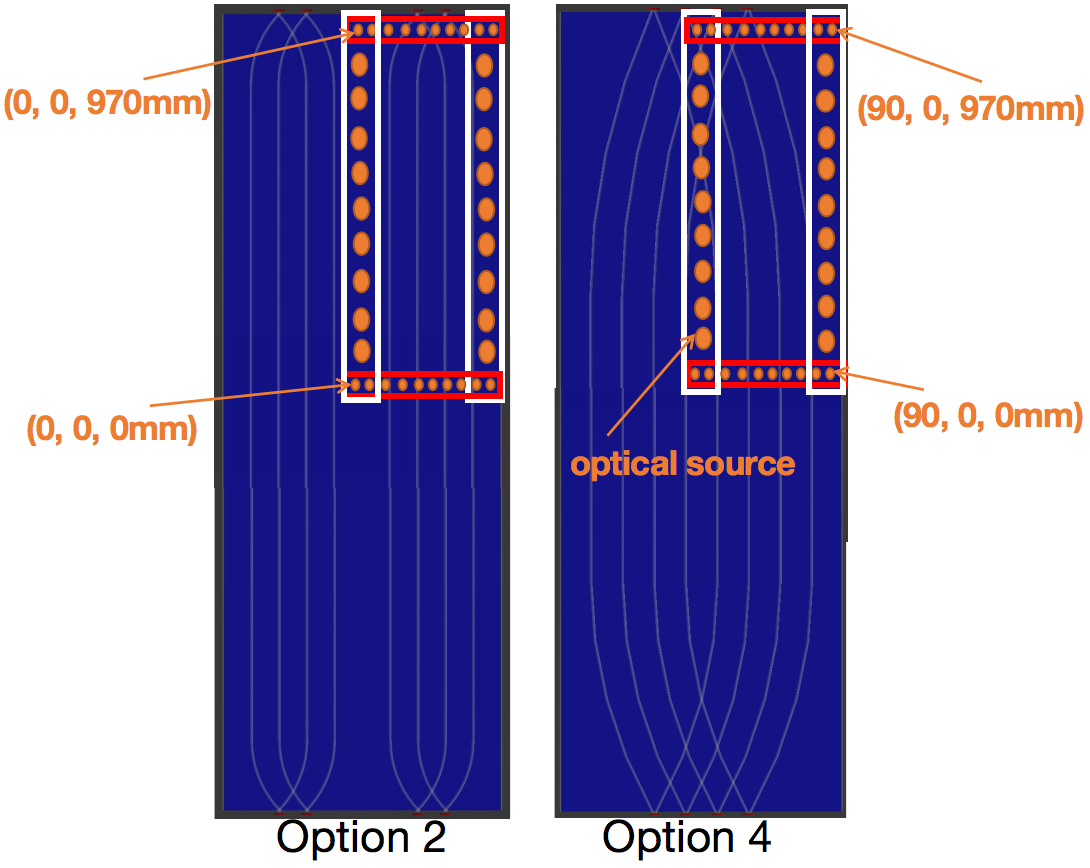}
\caption{Position distribution diagram of optical scanning of option 2 and option 4: the two white frames indicate the simulated points of 11 different positions along the length direction of the PS, respectively; and the two red frames indicate the simulated points of 10 different positions along the width direction of the PS, respectively. Each point is equidistant.}
\label{optS}
\end{center}
\end{figure}

Option 2 and option 4 have a similar configuration on WLS-fiber number, PS length, width, and thickness, but their light yields show an obvious difference. The photon transmission process is further checked for a better understanding of any specific reasons for the influence of fiber arrangement on the light yield for possible further optimization.

Muon deposits its energy along its track in the PS and emits photons in four $\pi$ directions. Meanwhile, the photons in the PS will propagate through attenuation, absorption, and re-emission effects, so it is not easy to accurately characterize the light transmission performance between different options. 
A specified optical survey is done to check the photon refection times before absorption by the WLS-fiber of each photon generated by PS, where the survey locations are shown in Figure \ref{optS} trying to cover the center and edge of the PS strip. The orange points are the specified locations to generate optical photons. At each point, 15000 photons are generated in four $\pi$ directions to mimic the random photons excited by a muon. 
It is obvious that before the photons enter the fiber, more reflection times, more difficult to reach SiPM (inverse relationship). At the same time, if the number of photons entering the fiber is more, the number of photons arriving at SiPM is more (proportional relationship). In order to reflect the combination of the two factors, an R-value proposed as the total number of photons entering the fiber divided by the average number of reflections of photons before entering the fiber.


\begin{figure*}[!htbp]
\begin{center}
\includegraphics[scale=0.4]{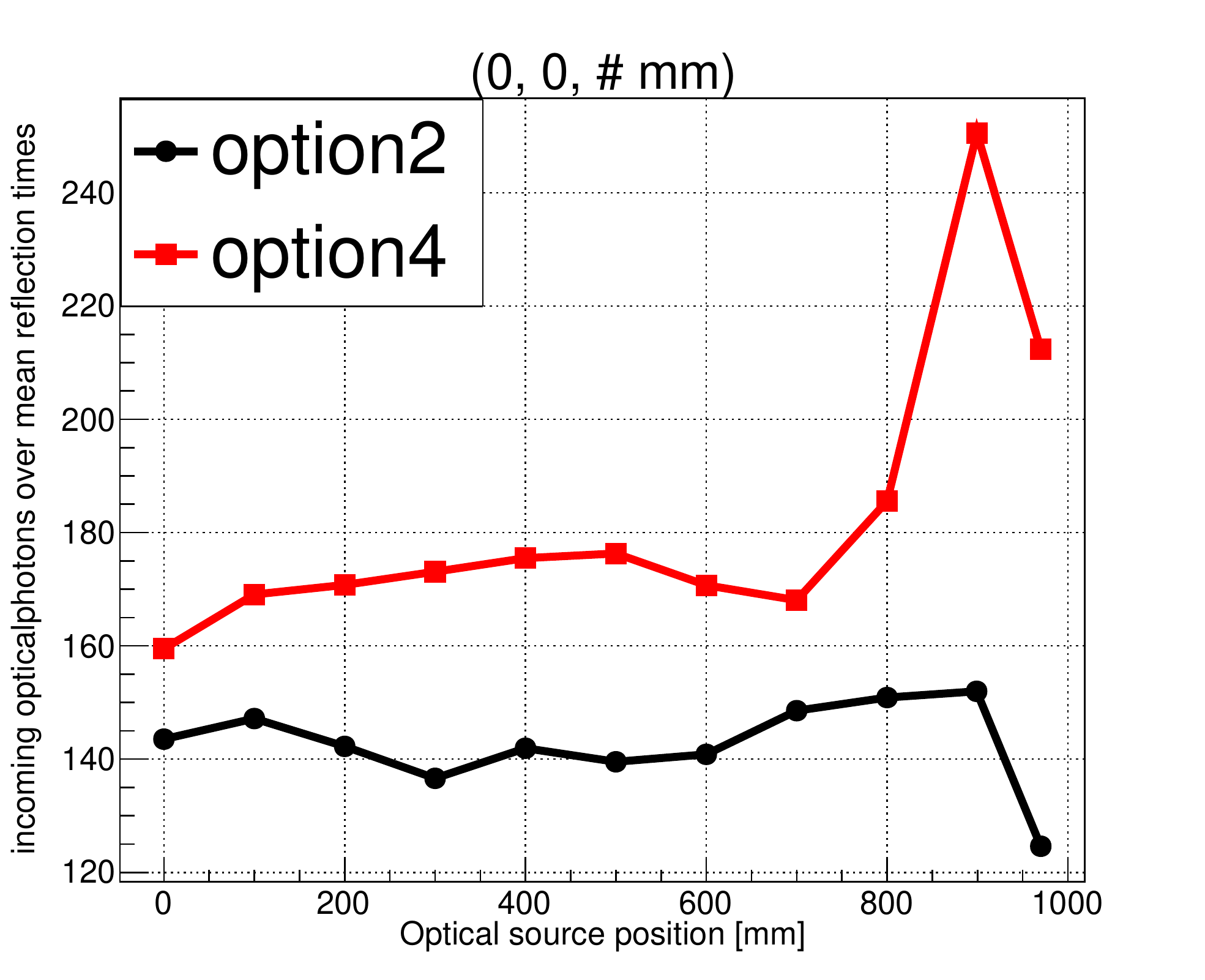}
\includegraphics[scale=0.4]{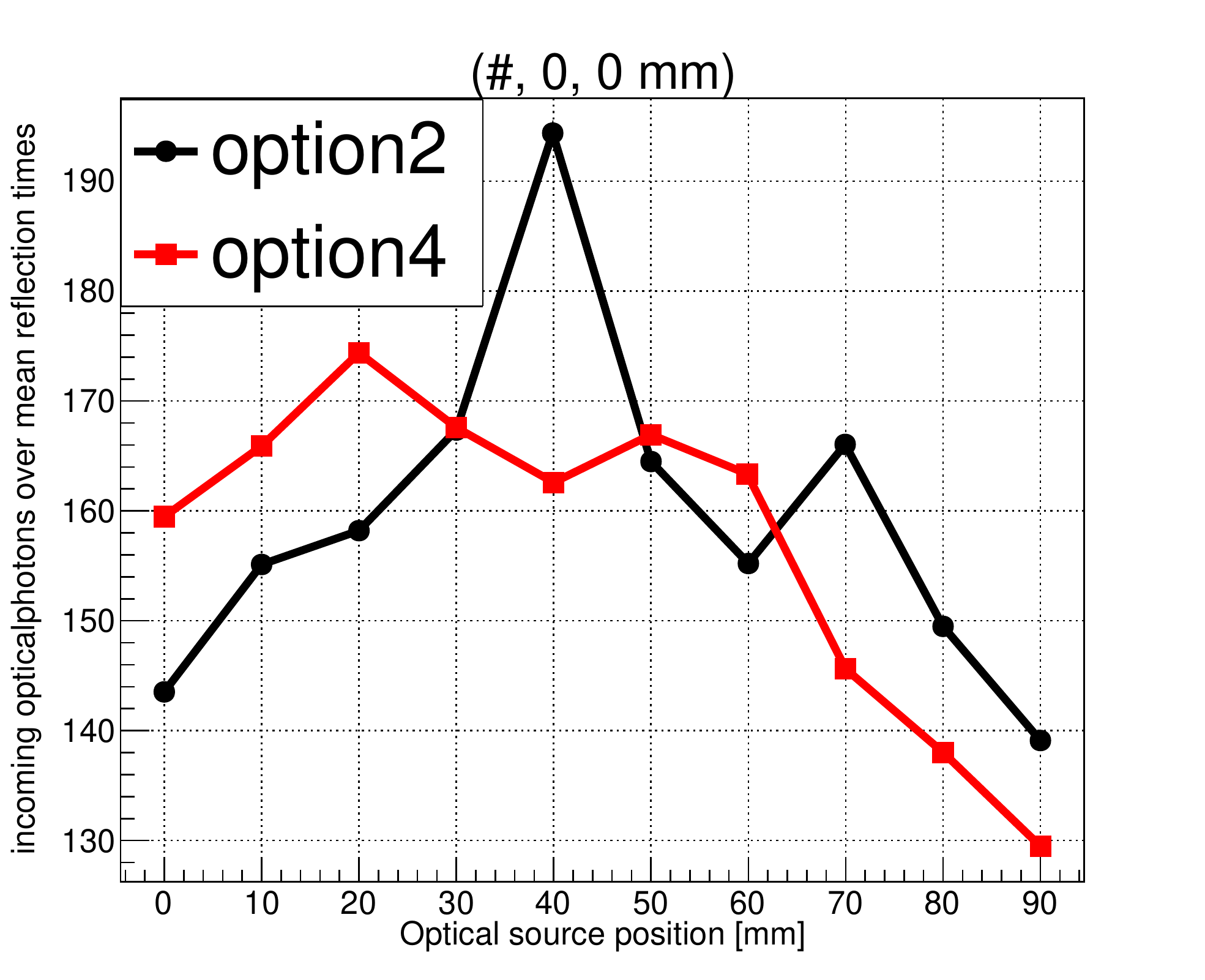}
\includegraphics[scale=0.4]{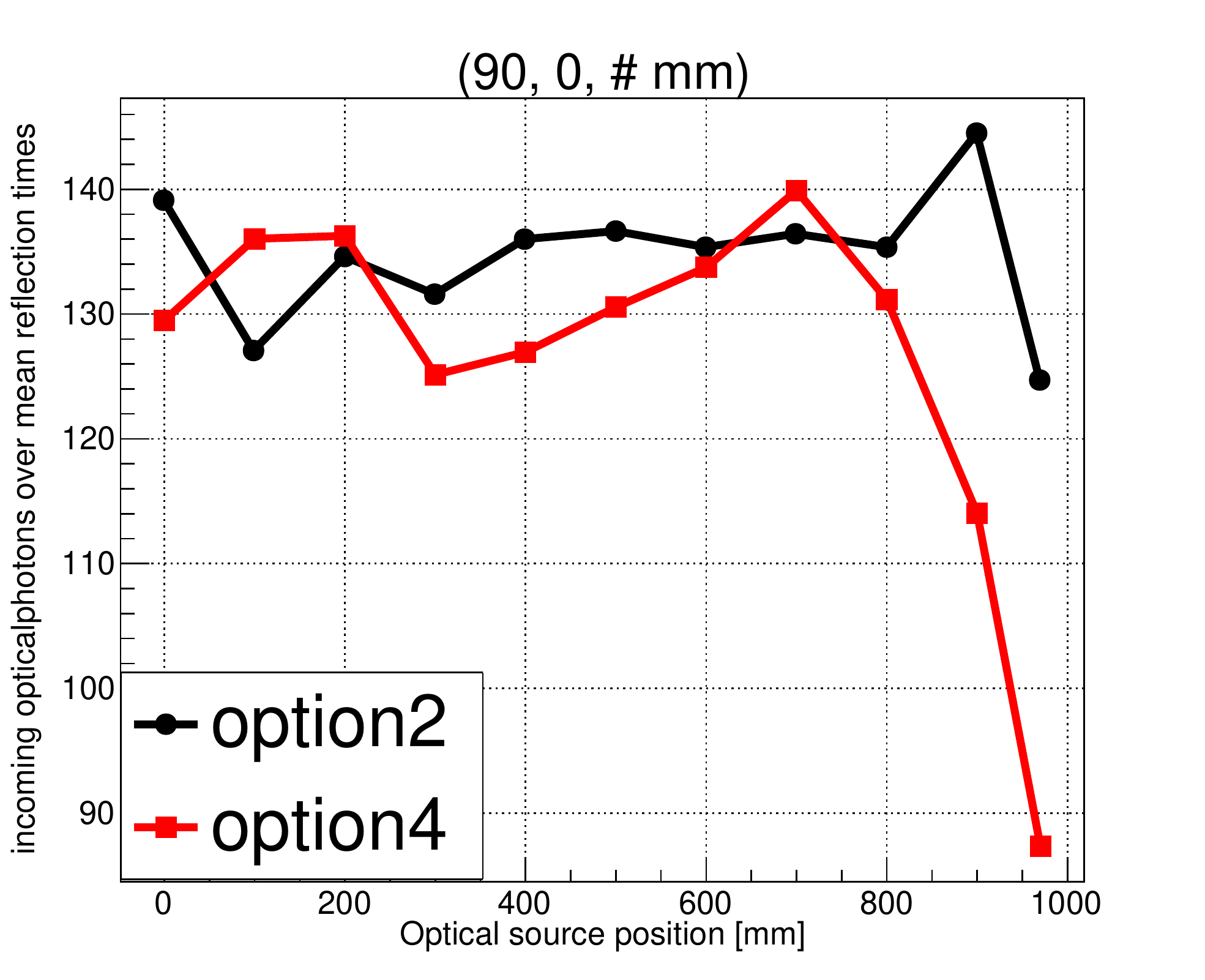}
\includegraphics[scale=0.4]{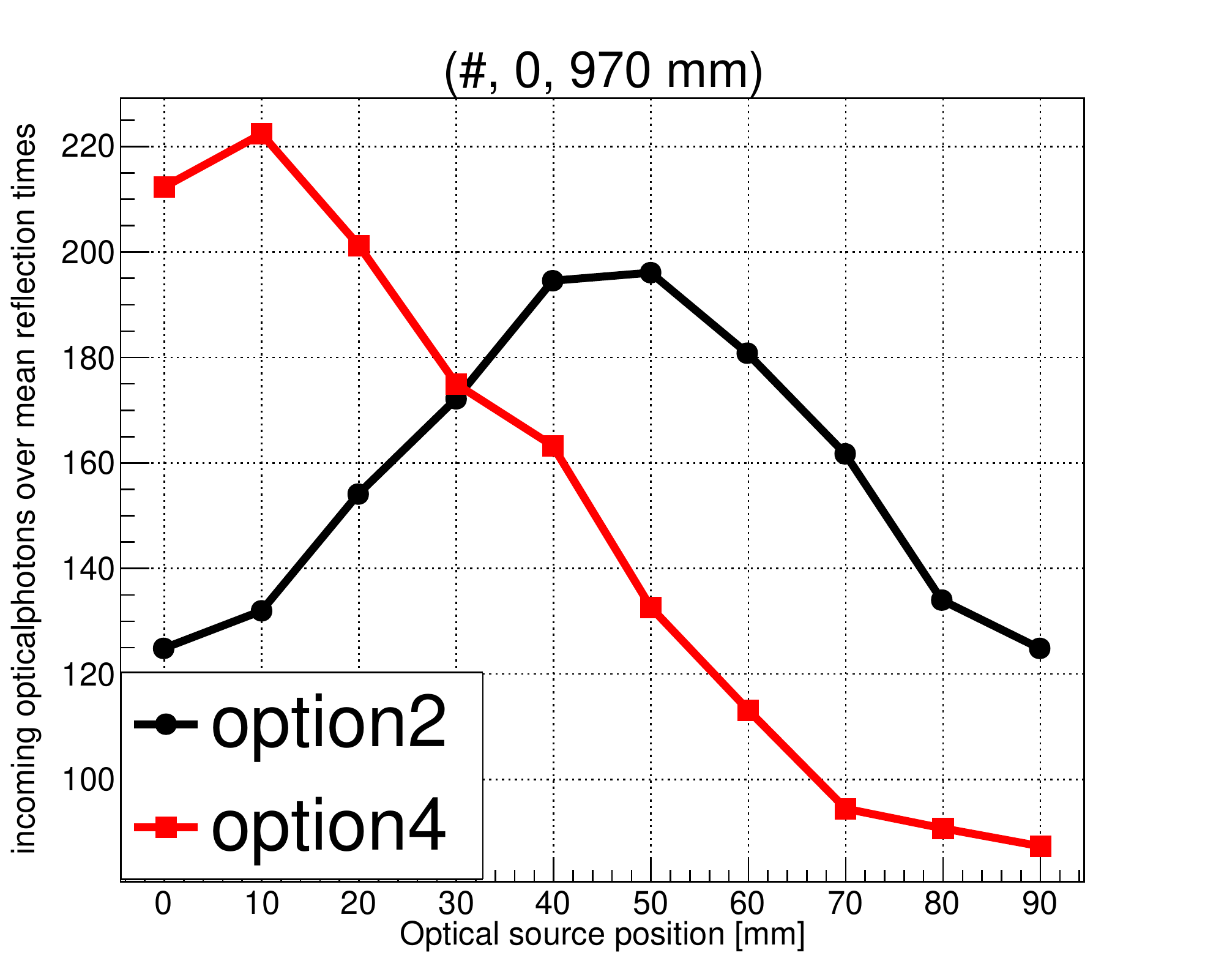}
\caption{The distribution of ratio of the photon numbers entering the fiber to the average times of reflections at different positions of the optical survey. The value of the Y coordinate of each point in the graph represents the R-value. The larger the R-value, the better the transmission performance. The results show that when the photon is generated at a location where the fiber is sparse, the number of photons entering the optical fiber will decrease sharply, while the times of reflections will also increase. Meanwhile, when photons is generated in dense places of fibers, the number of photons entering the fiber will increase dramatically, while the times of reflections will also decrease dramatically.We calculate the integral area under the red and the black line in the upper left and lower right panel of the Figure, then multiply the integral area of the red line in the upper left panel by the area in the lower right panel as the transmission performance of the whole region of option 4, do the same for the black as for the red line; The product of the two integral areas is used as the transmission performance of the entire region of option 2. Finally; We divide the product of the integral area of the red line by the product of the black, and find that the ratio is 2.21. This shows that the transmission performance of the whole area of option 4 is 2.21 times better than that of option 2.}
\label{24Sou}
\end{center}
\end{figure*}

Figure \ref{24Sou} shows distribution of R-value of optical survey at different positions under option 2 and option 4. The abscissa of each point is the position corresponding to the dash \# symbol in the figure, the specific location can be seen in combination with Figure \ref{optS}. The ordinate of each point is the R-value stands for optical transmission performance. The red line is the R-value of option 4, and the black represent option 2. The upper left panel of the figure shows the distribution of the R-value along the length direction of the PS strip when the position is the center of the width of the PS. The R-value of option 4 is higher than option 2 in the whole length direction. The lower left panel shows the distribution of the R-value along the length direction of the PS strip when the position in the edge of the width of the PS strip. The R-values of the two are nearly the same, except for the edge in the length direction, where the R-value of option 2 is greater than option 4. The upper right panel reveals the distribution of the R-value along the width direction of the PS when the position is the center of the length of the PS, where option 4 and option 2 have a trend of trade-offs. The lower right panel reveals the distribution of the R-value along the width direction of the PS when the position is the edge of the length direction of the PS, the R-value of option 4 is larger than that of option 2 at the positions within 30\,mm from the middle in the width direction of the PS. 

The results show that when the photon is generated at a location where the fiber is sparse, the number of photons entering the optical fiber will decrease sharply, while the times of reflections will also increase. Meanwhile, when photons is generated in dense places of fibers, the number of photons entering the fiber will increase dramatically, while the times of reflections will also decrease dramatically.

Looking at the four figures as a global picture, with option 2 configuration, the maximum vertical coordinate of the black point is less than 200, but with option 4 configuration, the ordinate of the red point can be greater than 200. 
The R-value of option 4 and option 2 has a trend of trade-offs and the overall difference is small from the lower left and upper right panel of Figure \ref{24Sou}. The R-value of option 4 and option 2 has a big differences from the upper left and lower right panel of Figure \ref{24Sou}. The distribution of R values in the PS region has been get, To obtain a quantitative relationship between the optical transmission performance of option 2 and option 4 in the entire region, An explanation of mathematical integration (R-value) were provided. We calculate the integral area under the red and the black line in the upper left and lower right panel of Figure \ref{24Sou}, then multiply the integral area of the red line in the upper left panel by the area in the lower right panel as the transmission performance of the whole region of option 4, do the same for the black as for the red line; The product of the two integral areas is used as the transmission performance of the entire region of option 2. Finally; We divide the product of the integral area of the red line by the product of the black, and find that the ratio is 2.21. This shows that the transmission performance of the whole area of option 4 is 2.21 times better than that of option 2.
Therefore, we can understand why the overall photon number of option 4 is more than that of option 2. At the same time, we can also see from the curve that the R-value distribution range of option 4 is larger than that of option 2, which means that the optical export uniformity of option 4 is worse than that of option 2, which is the result of uneven optical fiber layout.
In short, we found if the optical fiber in the PS middle region is in higher density and uniform, the higher the effective light yield will be, Therefore, to achieve a higher light yield, it is possible to consider placing as many optical fibers as possible in the middle of the PS.

So far, in terms of light yield, many simulations and made massive comparisons have done, The best configuration have obtained that is option 4 under the current design conditions. At the same time, to some extent, The reason why it is the best option have been quantitatively explained and a good reference and guidance suggestions for the processing technology and structure design were provided. 

For a PS detector, the most important factor, in addition to light yield, is the muon tagging efficiency. In the next section, The muon tagging efficiency and inefficiency are discussed.
\subsection{Muon tagging efficiency and inefficiency}
To simplify the simulation process, assuming that the photon detection efficiency (PDE) of SiPM is equal to 30\%. The sum of four SiPM outputs at both ends can be acquired under different options to further study the muon tagging efficiency. 
\begin{figure*}[!htbp]
\begin{center}
\includegraphics[scale=0.28]{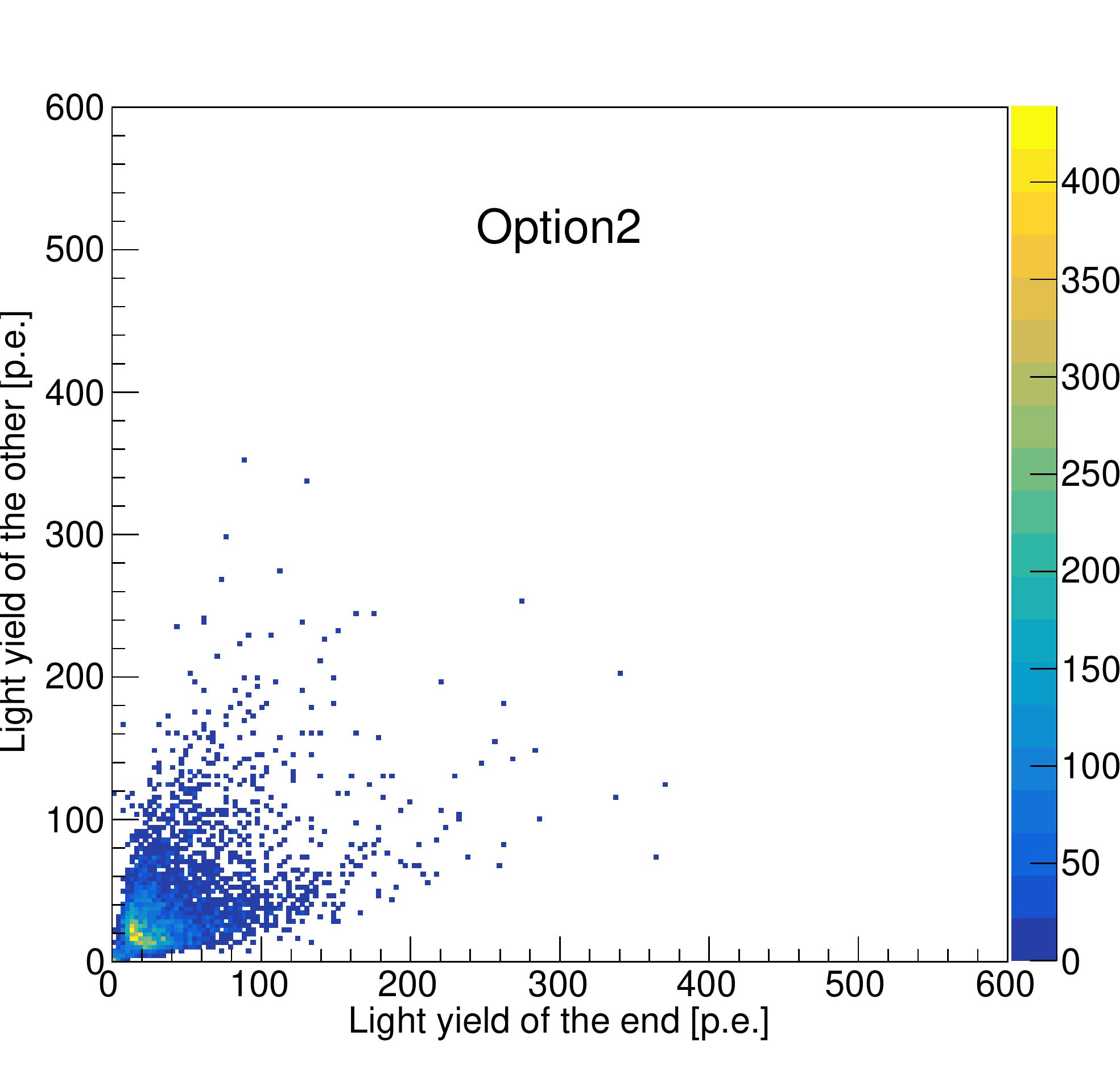}
\includegraphics[scale=0.28]{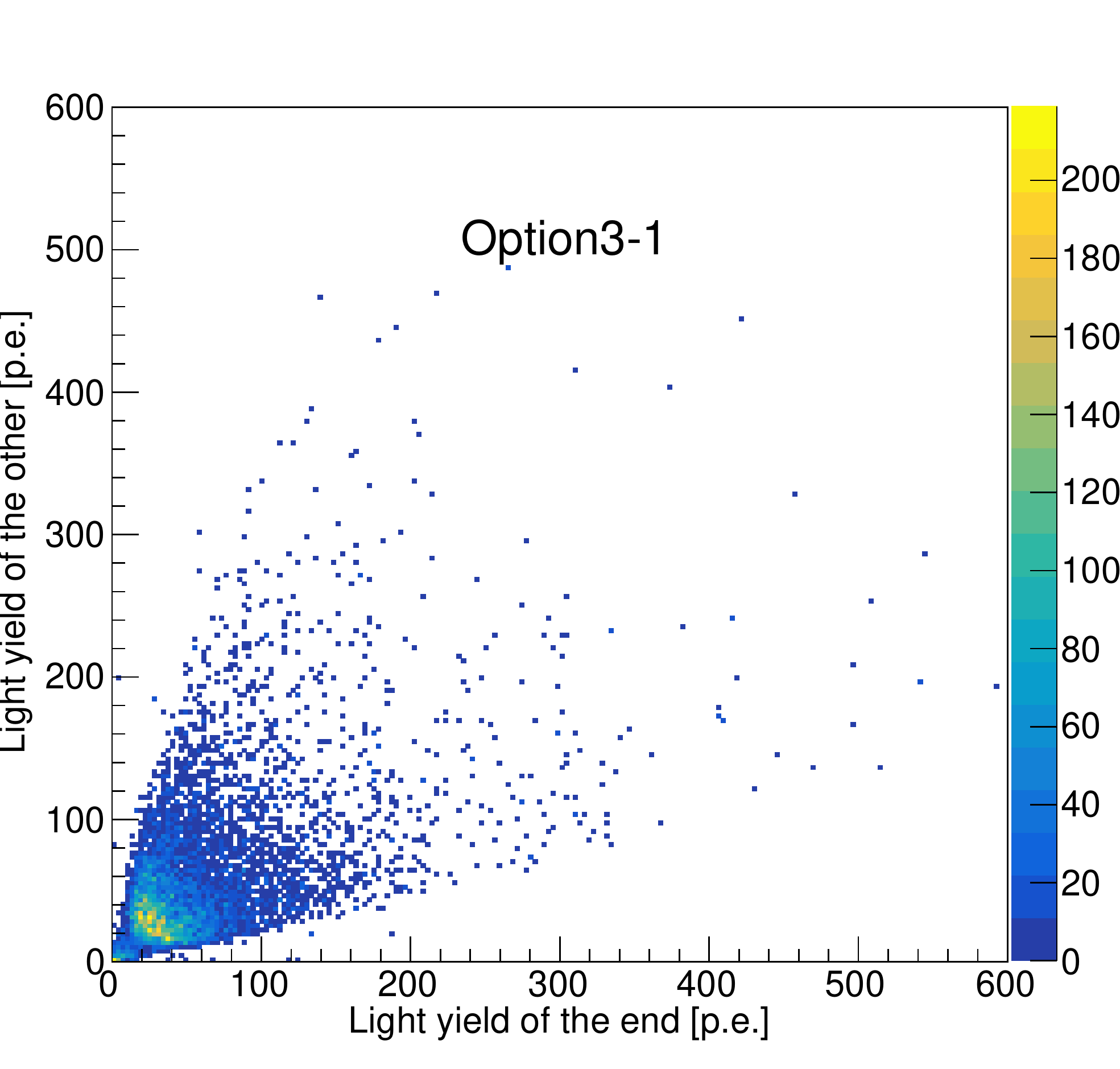}
\includegraphics[scale=0.28]{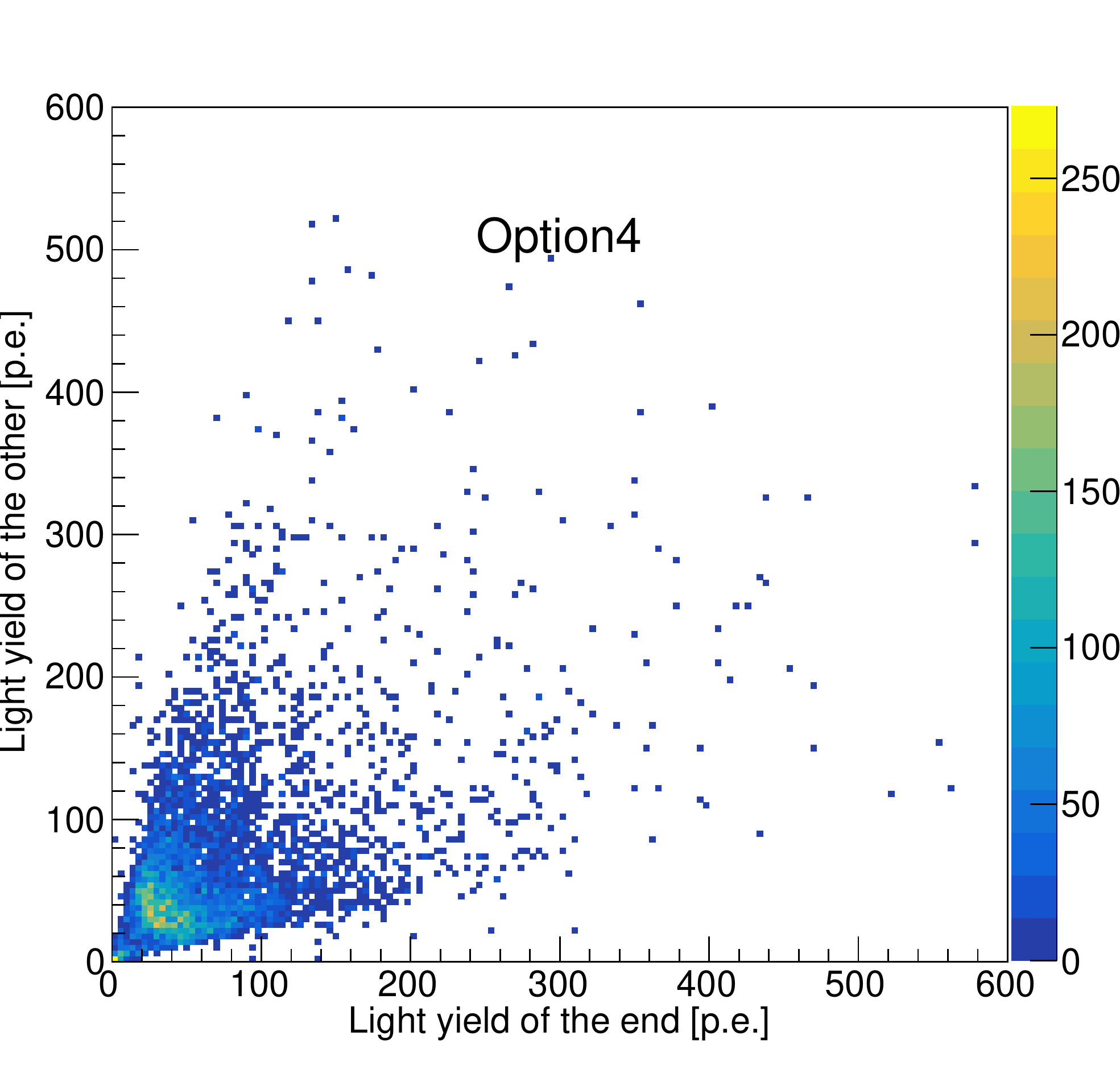}
\caption{Assuming 40\% SiPM PDE; From left to right: two-dimensional distribution of light yield at both ends of option 2, option 3-1 and option 4, respectively. The color represents the density of events.}
\label{234PE}
\end{center}
\end{figure*}
Figure \ref{234PE} shows the two-dimensional distribution of light yield at both ends of option 2, option 3-1, and option 4 in simulation, respectively. The abscissa is the sum of the SiPM outputs at one PS end, and the ordinate is the sum at the other end. The range of light yield at both ends of option 4 is the strongest and most divergent. If the sums at both ends of an event are greater than their respective threshold at the same time, it is considered that this muon has been detected. The event count of muon hitting the PS strip is noted as $N_{all}$, with the event count of detected muon noted as $N_{tag}$. The muon tagging efficiency is defined by the ratio $N_{tag}$/$N_{all}$. Here, the environmental background is not considered, mainly to eliminate environmental background interference.
\begin{figure*}[!htbp]
\begin{center}
\includegraphics[scale=0.29]{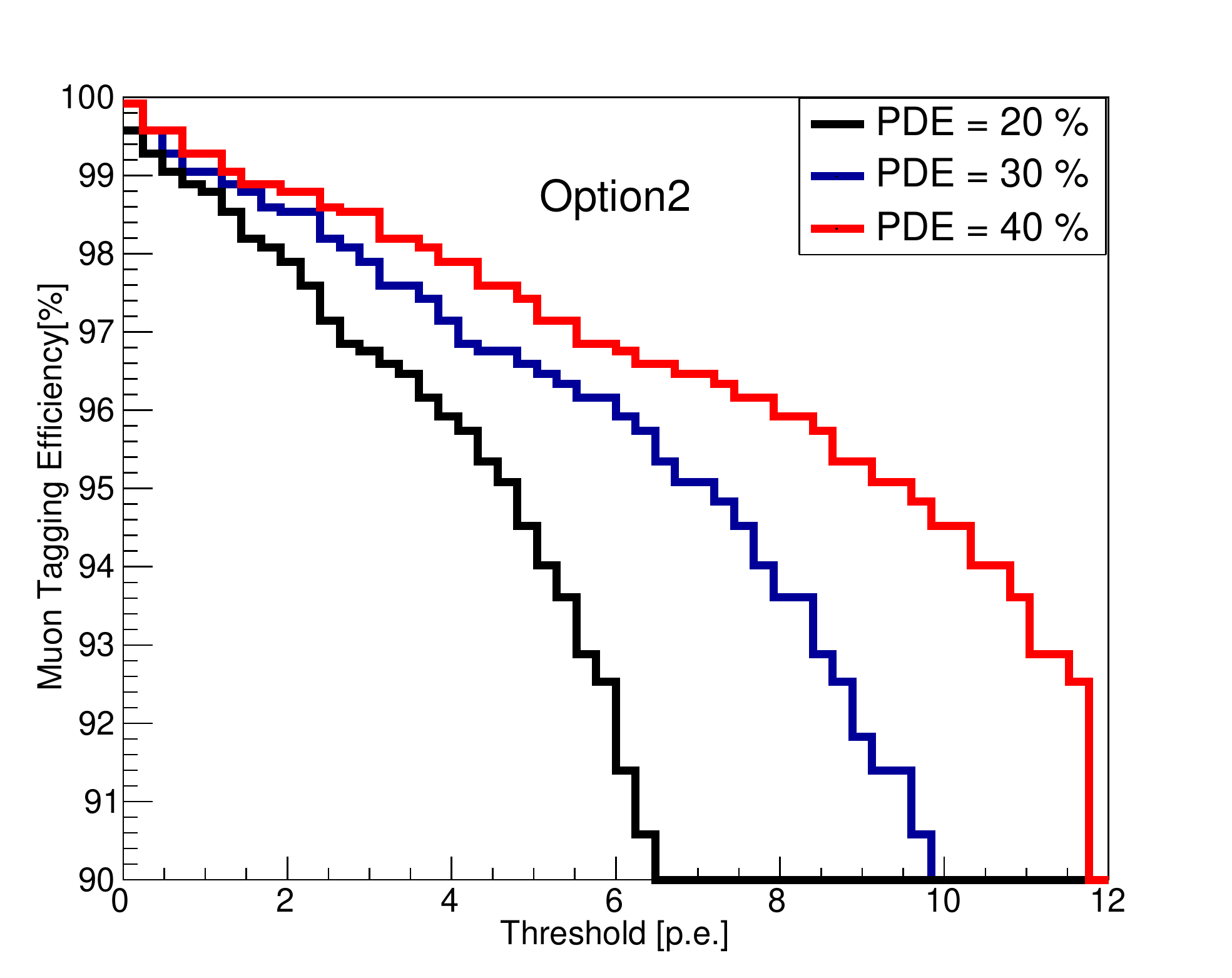}
\includegraphics[scale=0.29]{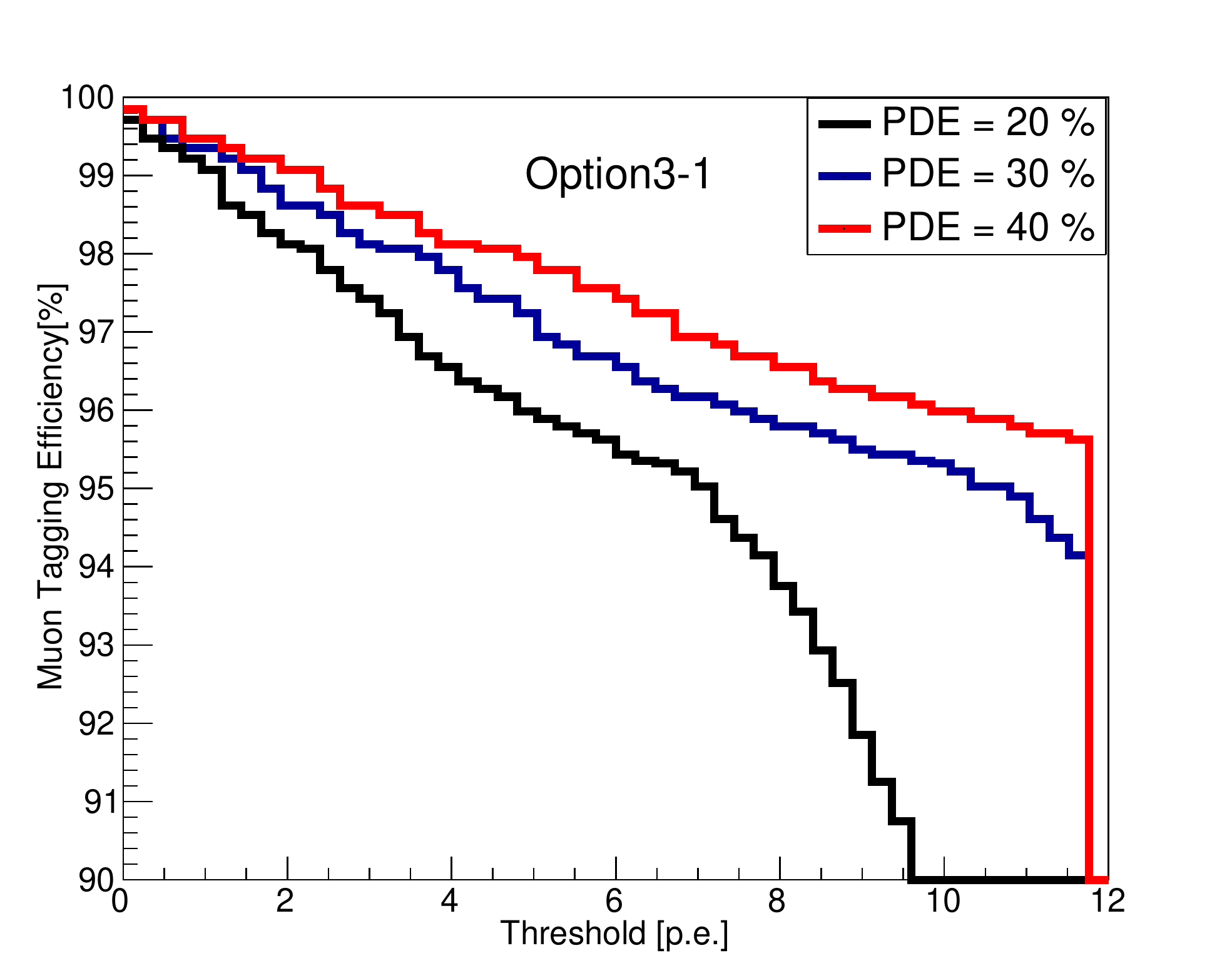}
\includegraphics[scale=0.29]{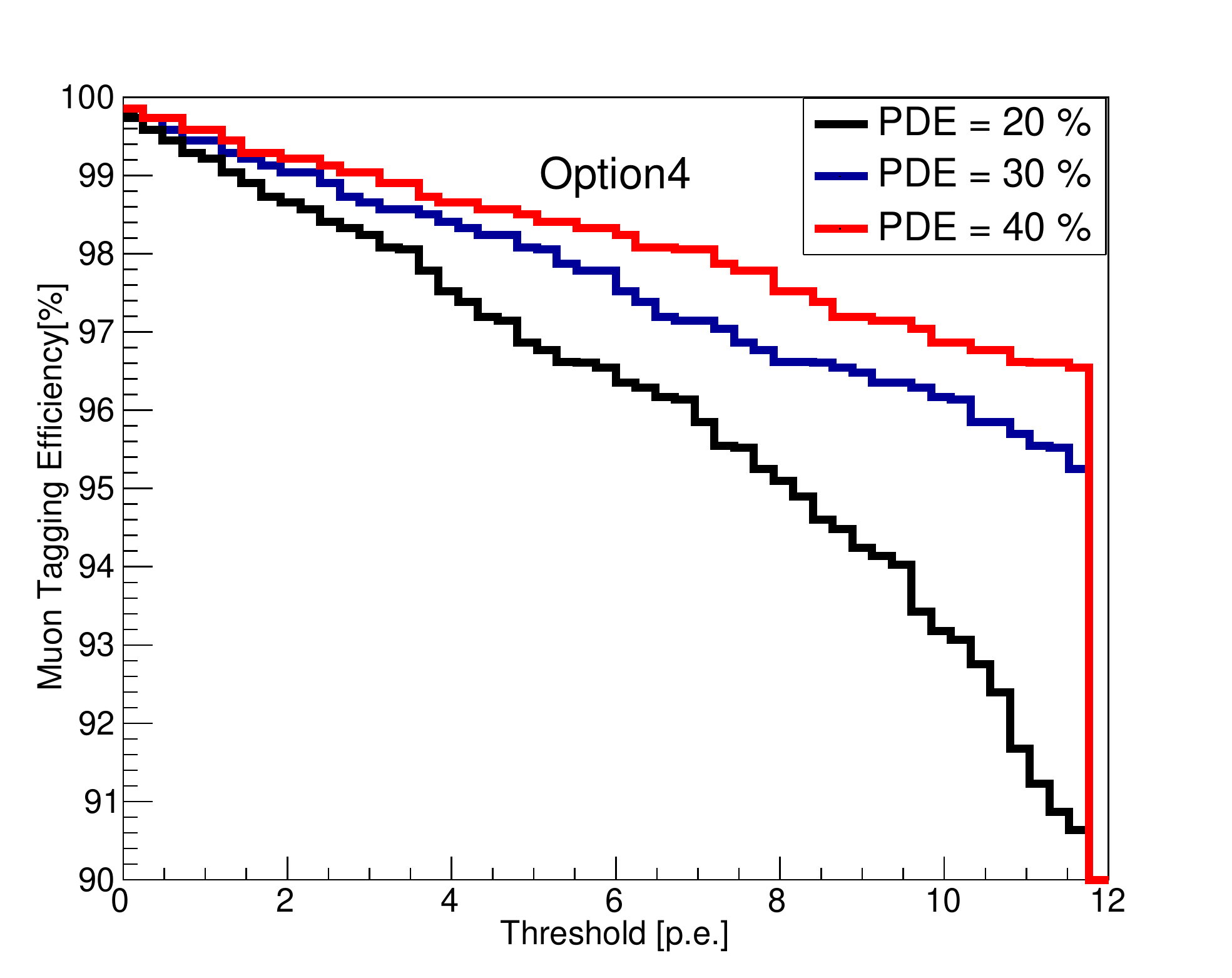}
\caption{Relationship between the muon tagging efficiency and the threshold of option 2, option 3-1, option 4, respectively. Where the threshold represents that the sum of the SiPM of each PS end have over threshold individually.}
\label{234VE}
\end{center}
\end{figure*}
Figure \ref{234VE} shows the relationship between the muon tagging efficiency and the threshold of option 2, option 3-1, and option 4, respectively. The black, blue and red represent the relationship between the efficiency and threshold when the PDE of SiPM is 20\%, 30\% and 40\%, respectively. As seen from the three figures, when PDE is 20\% and the threshold is set to 10\,p.e., the tagging efficiency of option 4 is still higher than 90\%, while option 2 and option 3-1 are both less than 90\%. With the increase of the threshold, the efficiency of option 2 decreases fastest, followed by option 3-1, and option 4 is the slowest. According to Figure \ref{234VE}, we can find the required threshold when we need to achieve certain tagging efficiency. 
\begin{figure}[!htbp]
\begin{center}
\includegraphics[scale=0.35]{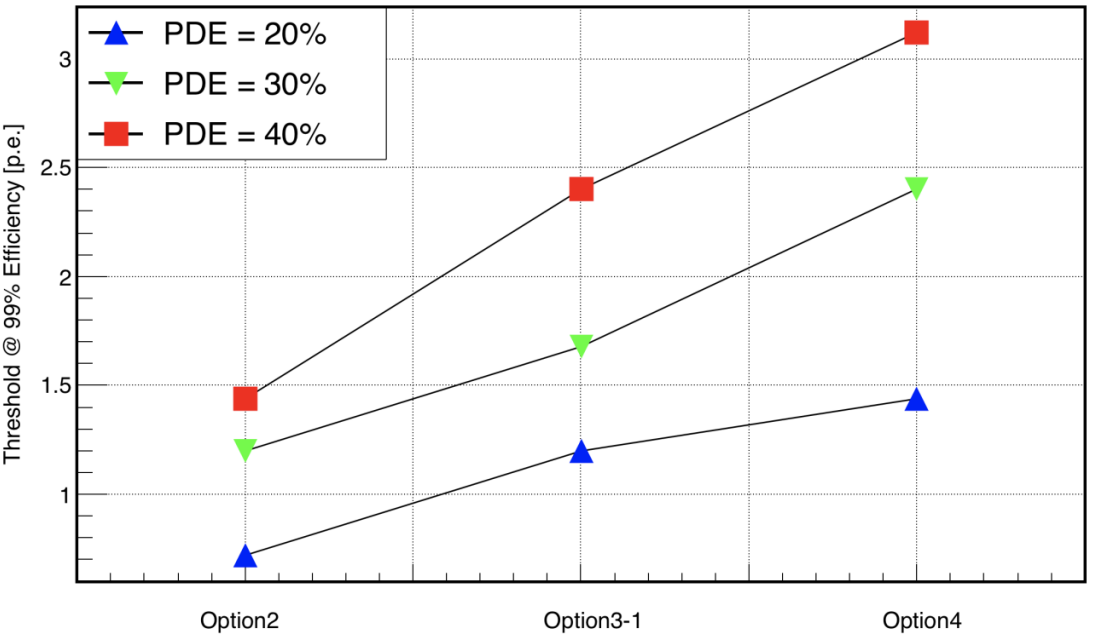}
\caption{With a muon tagging efficiency of 99\%, the threshold needed by option 2, option 3-1, and option 4, respectively.}
\label{234The}
\end{center}
\end{figure}

Figure \ref{234The} shows the corresponding threshold when the muon tagging efficiency reaches 99\% under different PDEs of SiPM. When the PDE of SiPM is 40\%, for option 2, the threshold cannot exceed 1.5\,p.e.; for option 3, it cannot exceed 2.4\,p.e.; however, for option 4, the threshold can be set to 3\,p.e., which greatly reduces the contribution from dark noise of SiPM. So far, option 4 is superior to other options in terms of light yield and muon tagging efficiency. Therefore, we take option 4 as our current optimal configuration. When the PDE of SiPM is 40\%, The threshold and corresponding tagging efficiency of a single PS strip were summarized in table \ref{table3}.

\begin{table}[!htbp]
 \centering
 \caption{Threshold and corresponding tagging efficiency of single PS strip with 40\% SiPM PDE}
 \label{table3}
 \begin{tabular}{|c|c|c|c|c|c|}
 \hline
  Threshold(p.e.) & 3.1 & 6.3 & 10 & 15 & 19  \\\hline
  Tagging efficiency  & 99\% & 98\% & 97\% & 96\% & 95\%\\\hline
\end{tabular} 
\end{table}

\begin{figure}[!htbp]
\begin{center}
\includegraphics[scale=0.47]{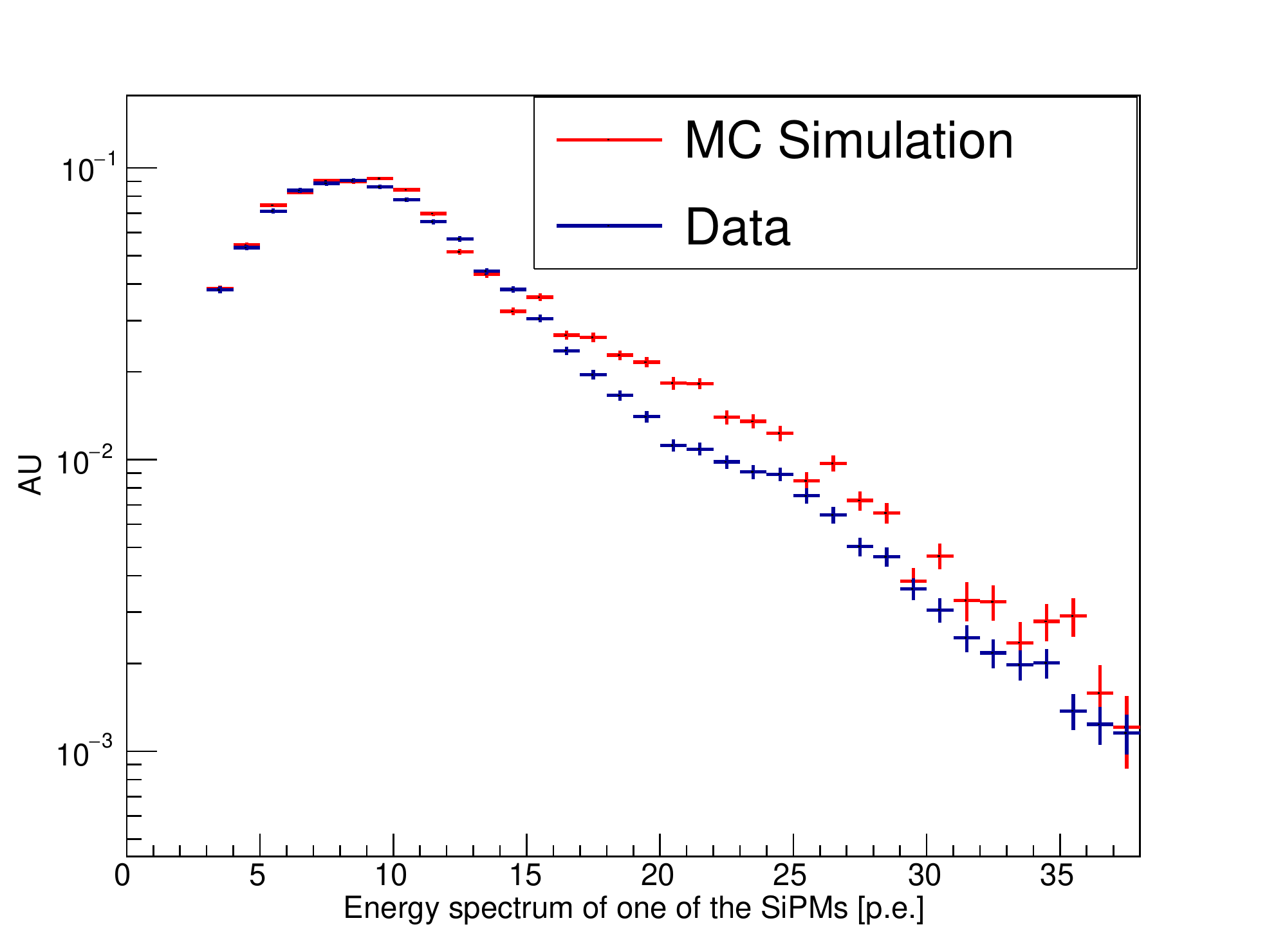}
\caption{Comparison between the simulation and experiment in terms of energy spectrum for option 4. Assuming 40\% SiPM PDE in MC. The PDE of SiPM sample provided by the manufacturer is 46\% at the wavelength of 420 nm when working voltage is 37.2 V. The data is recorded when the CR muon monitor in the center of PS.}
\label{MuonDatavsMC}
\end{center}
\end{figure}

To further verify the simulation, Another prototype with option 4 was build, the parameters of PS, optical fiber and reflective film are the same with before prototype in simulation. Only the detection efficiency of the backend SiPMs/PMT is different as labled. 2 SiPM were used as sensor and using oscilloscope to collect data. The SiPM is from the K-series MicroK-40035-TSV\cite{citeonsemi}. The data is saved by the lecroy-HDO4104A oscilloscope. Preliminary experimental results were obtained.

Figure \ref{MuonDatavsMC} shows the experimental measurement and simulation results of the PS prototype of option 4. To avoid the effect of the SiPM dark noise and environment background, Signals stronger than 3\,p.e. was analyzed, and the blue line is the energy spectrum of one SiPM from the measurement. The red line is the result of the simulation. Every point has an error bar, the abscissa is the light yield, and the ordinate is the log stand for the normalized event rate. It can be seen from the figure that in the spectrum below 12\,p.e., the simulated spectrum almost corresponds to the experiment. It was found that the most probable signal amplitude for passing through muons is around 8\,p.e. from data, So for the final configuration used for TAO where there are 4 SiPMs at each end of the PS strip, the sum of the most probable signal amplitude will be about 32\,p.e. When the spectrum is in the range of 15\,p.e.\,to 25\,p.e., within the range of error, the event rate of the simulated spectrum is more than that of the experiment. When the spectrum is greater than 25\,p.e.,the event rate of simulation and experiment events is relatively small. In the high energy region, the energy spectrum does not show good consistency, but it does not affect our optimization work.  There are several main factors that lead to differences in the energy spectrum. The first is that the coupling between SiPM and PS is not firm. The coupling method will be improve in the future. The second is that the widths of the two CRs are actually inconsistent, which may lead to a deviation of several centimeters in the location where Muon hits. Resulting in an impact on the energy spectrum.  

In any case, the light yield of option 4 is indeed much higher than that of option 1 before optimization. This shows that our optimization method is effective.
According to the above analysis and design, The light yield and the muon tagging efficiency corresponding to option 2, option 3-1, and option 4 have been obtained. When the muon tagging efficiency reaches 99\%, option 4 has the highest threshold limit. 

\begin{figure}[!htbp]
\begin{center}
\includegraphics[scale=0.47]{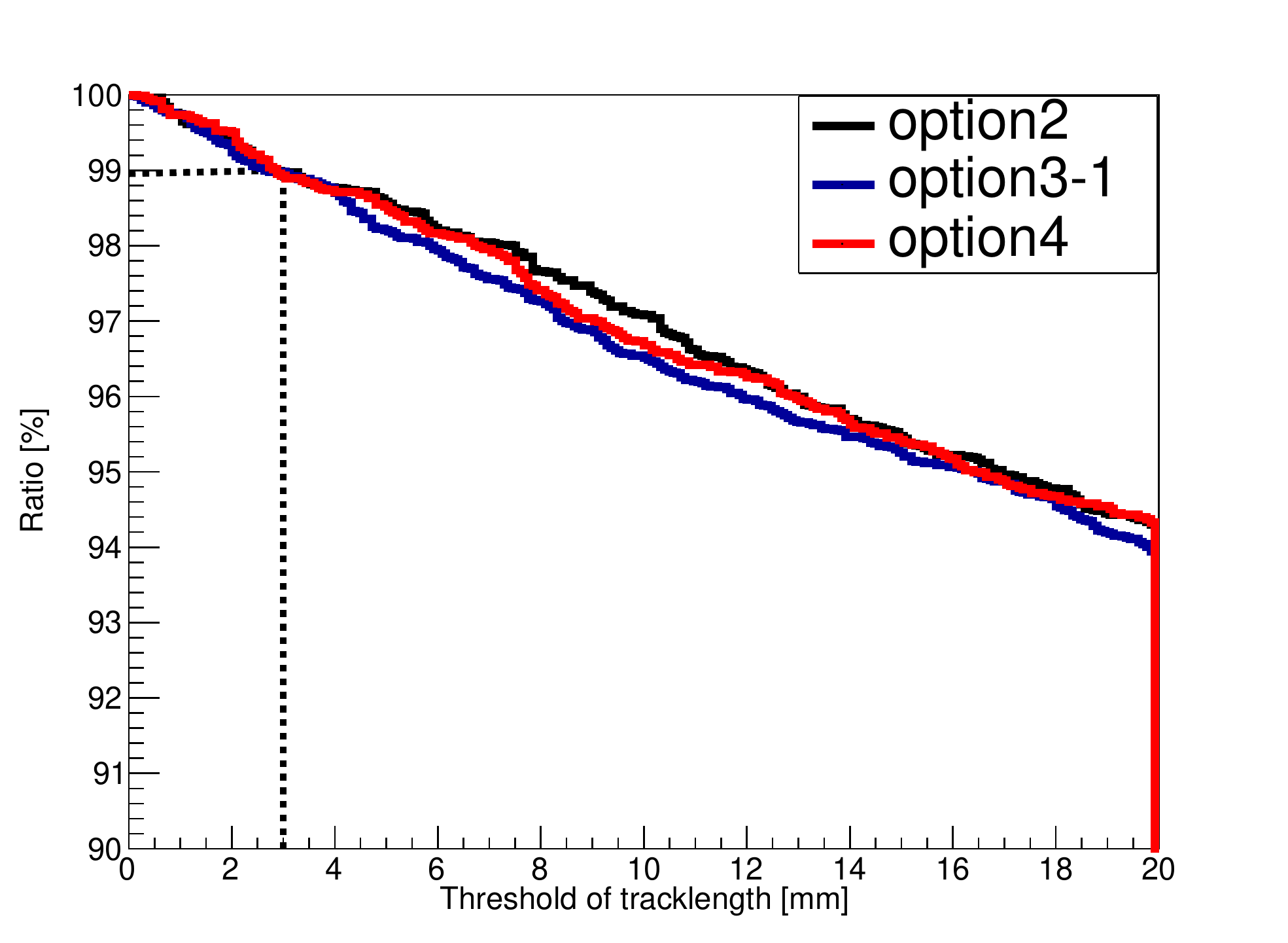}
\caption{Event ratio which the track length of incident muon in PS strip exceeds the threshold in the total events under option 2, option 3-1, and option 4, respectively.}
\label{234Tr}
\end{center}
\end{figure}

To study and discuss the inefficiencies by simulation, the track length of all muons in the PS have been counted. The inefficiencies can be indirectly explain by distribution of track length.

Figure \ref{234Tr} shows the ratio of the muon events whose track length exceeds certain threshold out of the total events under option 2, option 3-1, and option 4. 
The trends are almost the same and the difference in ratio was within 1\%. The muon generator, the thickness, length, and width of the PS strip are the same for the three configurations, and the declining trend of the event ratio is basically consistent under the three configurations.  
The black horizontal dotted line shows the proportion of 99\%, from which it can be obtained that the track length threshold of the corresponding black vertical dotted line is 3\,mm. There are 1\% muon events whose track length in the PS strip is less than 3\,mm. Since the thickness of the PS is 20\,mm, it is certain that the 1\% muon events are incident at a large zenith angle and pass through the edge of the PS, which is what we often call the edge event. Then, if building a module of PS to stagger the PS up, down, left and right, these edge events should also leave longer tracks on the next layer or the PS next to them, so as to be triggered. Thus, for a PS strip, the edge event may not be tagged(detected), but for the whole veto module system, it can be tagged. Therefore, under the same threshold, the muon tagging efficiency of a single layer module is higher than that of a single PS strip. 
Based on this point of view, the JUNO-TAO top veto system was designed, called top veto tracker (TVT).

\section{Performance of JUNO-TAO TVT}
\label{sec:TPS}
\begin{figure*}
   \subfigure[]{
   \begin{minipage}[]{0.45\linewidth}
   \centering
   \includegraphics[width=8cm]{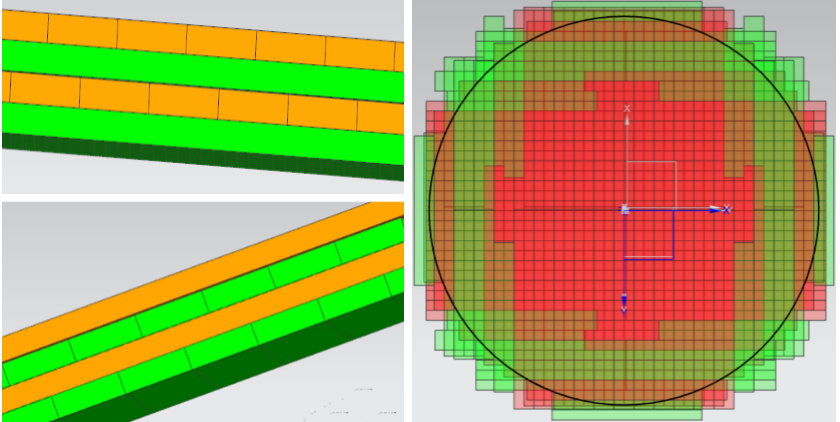}
    \label{5a}
    \end{minipage}
    }
    \subfigure[]{
    \begin{minipage}[d]{0.45\linewidth}
    \centering
    \includegraphics[width=9cm]{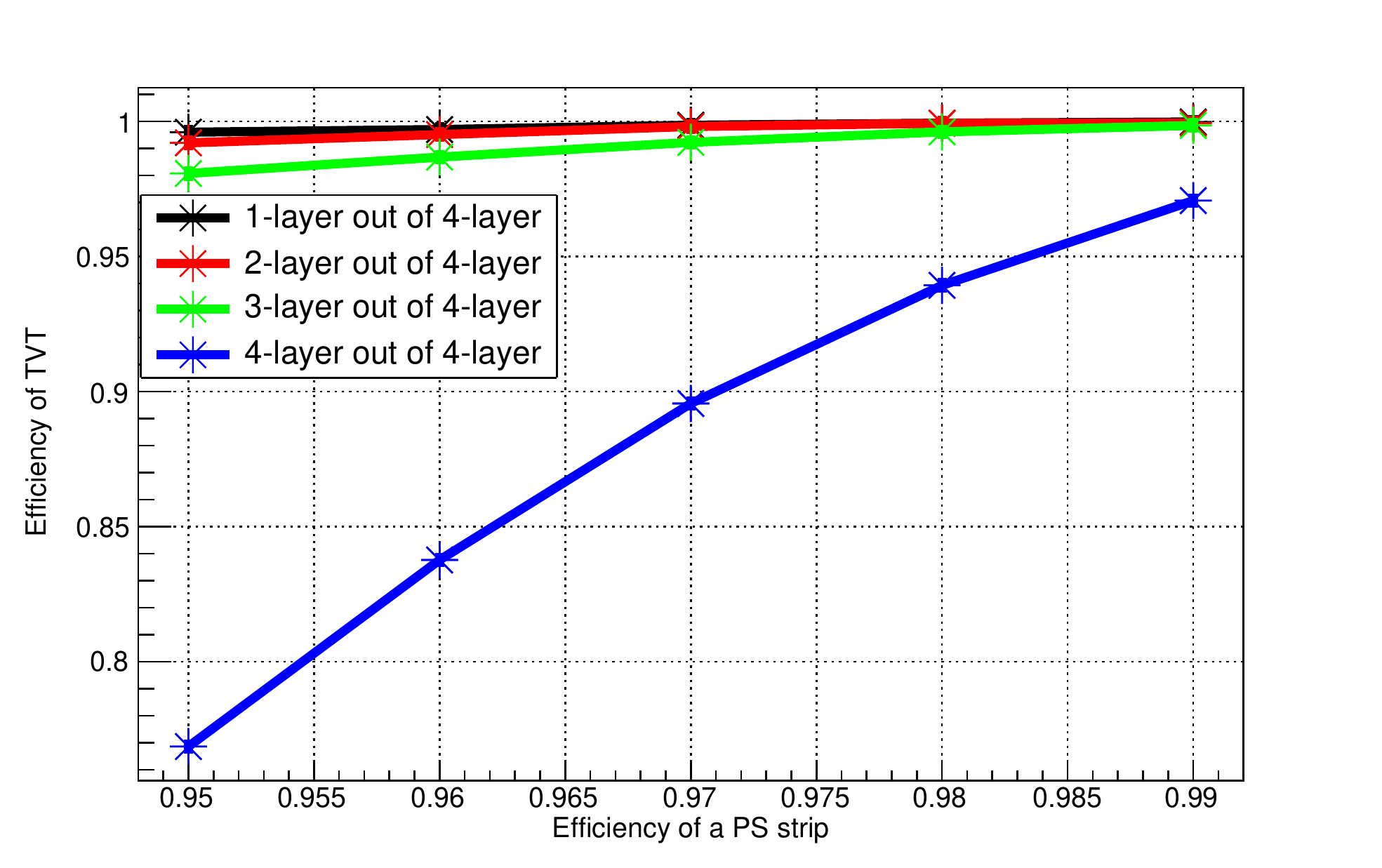}
    \label{5b}
    \end{minipage}
    }
    \caption{(a):Design of JUNO-TAO TVT. The left panel is a side view. The right panel is a top view.(b):Muon tagging efficiency of hitting multi-layer PS with that of single.}
    \label{TAO_tvt}
\end{figure*}
Figure \ref{5a} is the design schema of the TVT. The left panel of Figure \ref{5a} shows the side view of the whole 4-layer PS, the seam of the upper layer of PS corresponds to the volume of the next layer of PS, which can eliminate the dead space. At the same time, we consider that this system may be able to determine the track direction of a muon, so the current design is a 4-layer PS. The gap between neighboring layers is 2\,cm (It refers to the gap between the SiPM end of one PS and the end of another PS). The right panel of Figure \ref{5a} shows the top view of the whole 4-layer PS. Different colors represent different layers of PS. Since the central detector of TAO is a ball, under the condition of saving funds and consumables, we try our best to cover the projection surface of the ball with the area of PS. So the four layers form a circle. 

Figure \ref{5b} shows the distribution of the muon tagging efficiency of hitting multi-layer PS versus that of a single PS. The different colors represent the efficiencies of hitting different number of layers of PS. The black line represents the muon tagging efficiency of any one out of four layers. The red represents any two out of four layers. Because the gap between each layer is very small, the red and black lines almost coincide a PS strip efficiency. The blue represents the muon that hits the whole four layers and is also tagged by the four layers of PS. Since each layer has a certain dead space, the decrease of efficiency of tagging on all four layers will decrease most obviously with the efficiency of a PS strip. The green represents any three out of four layers, from which this conclusion can be obtained: even if the efficiency of a PS strip is as low as 97\%, which is equivalent to the threshold of 10\,p.e. can be found in Table \ref{table3}, the muon tagging efficiency of any three out of four layers will still be higher than 99\%. In addition, by simulating the environment background (including $^{238}U$, $^{232}Th$, $^{40}K$ chain), the background event rate that can pass 3 \,p.e. threshold in any two out of four layers of PS is 1280 $\pm$ 40 \,Hz and the rate that can pass 3 \,p.e. threshold in any three out of four layers of PS is < 10 \,Hz. Therefore, we can select the condition that any three out of the four layers are triggered to tag muon and reject the background.

\section{summary}
\label{sec:Sum}
There are numerous PS with WLS-fiber options with different configurations. 
In this paper, through simulation package, the differences between different configurations are checked in detail, for example, for the influence of optical fiber diameter, layout and other factors on transmission performance. Based on light yield and muon tagging efficiency, the optimal configuration of WLS-fiber PS under certain conditions is obtained.  
The relevant bench test of option 1 and option 4 has been measured, in terms of the most probable SiPM output, the summed signal at one end of the optimized option 4 is about 32\, p.e.. The light yield is 4 times larger than option 1, which further verified the reliability of the simulation package.

At the same time, inefficiency muon events are also studied. Finally, the design scheme of JUNO-TAO TVT system is reached. These conclusions can be given in the simulation: Assuming 40\% SiPM PDE, the muon tagging efficiency of a single PS strip at 3\,p.e. threshold can still reach 99\% in "AND" mode(i.e., when signals above the threshold was required from two sides of a PS strip.), the background event rate that can pass the threshold in any three out of four layers of PS is less than 10 Hz(Almost zero). When the threshold is 10\,p.e., the muon tagging efficiency of a single PS strip is 97\%, The efficiency of 3-layer out of 4-layer of TVT will still be higher than 99\%.

\bibliographystyle{unsrt}
\bibliography{references}

\begin{thebibliography}{10}

\bibitem{muon-rate}
P.A.~Zyla et~al. (Particle Data~Group).
\newblock {Review of Particle Physics}.
\newblock {\em Progress of Theoretical and Experimental Physics}, 2020, 08
  2020.
\newblock 083C01.

\bibitem{muon-energy}
Mengyun Guan, Ming-Chung Chu, Jun Cao, Kam-Biu Luk, and Changgen Yang.
\newblock A parametrization of the cosmic-ray muon flux at sea-level.
\newblock {\em Chinese Physics C}, 2015.

\bibitem{Patrignani:2016xqp}
C.~Patrignani et~al.
\newblock {Review of Particle Physics}.
\newblock {\em Chin. Phys. C}, 40(10):100001, 2016.

\bibitem{Guo_2021}
Zi~yi~Guo, Lars Bathe-Peters, Shao min Chen, Mourad Chouaki, Wei Dou, Lei Guo,
  Ghulam Hussain, Jin jing Li, Qian Liu, Guang Luo, Wen tai Luo, Ming Qi, Wen
  hui Shao, Jian Tang, Lin yan Wan, Zhe Wang, Ben da~Xu, Tong Xu, Wei ran Xu,
  Yu~zi~Yang, Minfang Yeh, Lin Zhao, and (JNE Collaboration).
\newblock Muon flux measurement at china jinping underground laboratory *.
\newblock {\em Chinese Physics C}, 45(2):025001, feb 2021.

\bibitem{BARBUTO2004485}
E.~Barbuto, C.~Bozza, M.~Cozzi, N.~D'Ambrosio, G.~{De Lellis}, G.~{De Rosa},
  M.~{De Serio}, L.S. Esposito, G.~Giacomelli, M.~Giorgini, G.~Grella, M.~Ieva,
  G.~Mandrioli, P.~Migliozzi, M.T. Muciaccia, L.~Patrizii, C.~Pistillo,
  P.~Righini, G.~Rosa, S.~Simone, M.~Sioli, C.~Sirignano, G.~Sirri,
  G.~Sorrentino, and V.~Tioukov.
\newblock Atmospheric muon flux measurements at the external site of the gran
  sasso lab.
\newblock {\em Nuclear Instruments and Methods in Physics Research Section A:
  Accelerators, Spectrometers, Detectors and Associated Equipment},
  525(3):485--495, 2004.

\bibitem{Trzaska2019}
Wladyslaw~Henryk Trzaska, Maciej Slupecki, Iulian Bandac, Alberto Bayo,
  Alessandro Bettini, Leonid Bezrukov, Timo Enqvist, Almaz Fazliakhmetov, Aldo
  Ianni, Lev Inzhechik, Jari Joutsenvaara, Pasi Kuusiniemi, Kai Loo, Bayarto
  Lubsandorzhiev, Alexander Nozik, Carlos Peña~Garay, and Maria Poliakova.
\newblock Cosmic-ray muon flux at canfranc underground laboratory.
\newblock {\em European Physical Journal C}, 79(8), 2019.
\newblock Cited by: 6; All Open Access, Gold Open Access, Green Open Access.

\bibitem{JUNO-detector}
{JUNO Collaboration}.
\newblock {JUNO} physics and detector.
\newblock {\em Progress in Particle and Nuclear Physics}, 123:103927, 2022.

\bibitem{JUNO-CDR}
{JUNO Collaboration} and T.~Adam et~al.
\newblock {{JUNO} Conceptual Design Report}.
\newblock {\em arXiv e-prints}, page arXiv:1508.07166, August 2015.

\bibitem{TAO-CDR}
{JUNO Collaboration} and Angel~Abusleme et~al.
\newblock {{TAO} Conceptual Design Report: A Precision Measurement of the
  Reactor Antineutrino Spectrum with Sub-percent Energy Resolution}.
\newblock {\em arXiv e-prints}, page arXiv:2005.08745, May 2020.

\bibitem{XENON1T:2014eqx}
E.~Aprile et~al.
\newblock {Conceptual design and simulation of a water Cherenkov muon veto for
  the XENON1T experiment}.
\newblock {\em JINST}, 9:P11006, 2014.

\bibitem{XENON:2020kmp}
E.~Aprile et~al.
\newblock {Projected WIMP sensitivity of the XENONnT dark matter experiment}.
\newblock {\em JCAP}, 11:031, 2020.

\bibitem{MAGIX:2022fdt}
Mirco Christmann et~al.
\newblock {Light Dark Matter Searches with DarkMESA}.
\newblock {\em PoS}, EPS-HEP2021:129, 2022.

\bibitem{DarkSide:2013syn}
T.~Alexander et~al.
\newblock {DarkSide search for dark matter}.
\newblock {\em JINST}, 8:C11021, 2013.

\bibitem{Pocar:2015ota}
Andrea Pocar.
\newblock {Searching for neutrino-less double beta decay with EXO-200 and
  nEXO}.
\newblock {\em Nucl. Part. Phys. Proc.}, 265-266:42--44, 2015.

\bibitem{Tosi:2013bta}
D.~Tosi.
\newblock {Search for double beta decay with EXO-200}.
\newblock {\em AIP Conf. Proc.}, 1560(1):187--189, 2013.

\bibitem{Gornea:2009zz}
Razvan Gornea.
\newblock {Double beta decay in liquid xenon}.
\newblock {\em J. Phys. Conf. Ser.}, 179:012004, 2009.

\bibitem{Birks:1964zz}
John~B. Birks.
\newblock {\em {The Theory and practice of scintillation counting}}.
\newblock 1964.

\bibitem{Zhezher:2020qov}
Y.~Zhezher.
\newblock {Study of Muons in Ultra-High-Energy Cosmic-Ray Air Showers with the
  Telescope Array Experiment}.
\newblock {\em Phys. Atom. Nucl.}, 82(6):685--688, 2020.

\bibitem{NUCLEUS:2022vyj}
Andreas Erhart, Victoria Wagner, Ludwig Klinkenberg, Thierry Lasserre, David
  Lhuillier, Claudia Nones, Rudolph Rogly, Vladimir Savu, R.~Strauss, and
  Matthieu Vivier.
\newblock {Development of an Organic Plastic Scintillator based Muon Veto
  Operating at Sub-Kelvin Temperatures for the NUCLEUS Experiment}.
\newblock In {\em {19th International Workshop on Low Temperature Detectors}},
  5 2022.

\bibitem{Seo:2022vzr}
J.~W. Seo, E.~J. Jeon, W.~T. Kim, Y.~D. Kim, H.~Y. Lee, J.~Lee, M.~H. Lee,
  P.~B. Nyanda, and E.~S. Yi.
\newblock {A feasibility study of extruded plastic scintillator embedding WLS
  fiber for AMoRE-II muon veto}.
\newblock {\em Nucl. Instrum. Meth. A}, 1039:167123, 2022.

\bibitem{veto-LBNL-THOMAS201347}
K.J. Thomas, E.B. Norman, A.R. Smith, and Y.D. Chan.
\newblock Installation of a muon veto for low background gamma spectroscopy at
  the lbnl low-background facility.
\newblock {\em Nuclear Instruments and Methods in Physics Research Section A:
  Accelerators, Spectrometers, Detectors and Associated Equipment}, 724:47--53,
  2013.

\bibitem{Pla-Dalmau:2000puk}
A.~Pla-Dalmau, A.~D. Bross, and K.~L. Mellott.
\newblock {Low-cost extruded plastic scintillator}.
\newblock {\em Nucl. Instrum. Meth. A}, 466:482--491, 2001.

\bibitem{Moiseev:2007zz}
A.~A. Moiseev, R.~C. Hartman, T.~E. Johnson, D.~J. Thompson, P.~L. Deering,
  T.~R. Nebel, and J.~F. Ormes.
\newblock {High Efficiency Plastic Scintillator Detector with
  Wavelength-Shifting Fiber Readout for the GLAST Large Area Telescope}.
\newblock {\em Nucl. Instrum. Meth. A}, 583:372--381, 2007.

\bibitem{Vaishali2021Design}
Vaishali~Manojkumar Thakur, Amit Jain, P.~Ashokkumar, Rekha Anilkumar, Pravin
  Sawant, Probal Chaudhury, and L.M. Chaudhari.
\newblock Design and development of a plastic scintillator based whole body
  beta/gamma contamination monitoring system.
\newblock {\em Nuclear Science and Techniques}, 32(5), 2021.

\bibitem{Holm:1989sb}
U.~Holm and K.~Wick.
\newblock {Radiation Stability of Plastic Scintillators and Wave Length
  Shifters}.
\newblock {\em IEEE Trans. Nucl. Sci.}, 36:579--583, 1989.

\bibitem{Bloise:2023xhc}
C.~Bloise et~al.
\newblock {Design, assembly and operation of a Cosmic Ray Tagger based on
  scintillators and SiPMs}.
\newblock {\em Nucl. Instrum. Meth. A}, 1045:167538, 2023.

\bibitem{Buzhan:2020ryp}
P.~Buzhan and A.~Karakash.
\newblock {Hand-foot monitors for nuclear plants based on
  scintillator\textendash{}WLS\textendash{}SiPM technology}.
\newblock {\em J. Phys. Conf. Ser.}, 1689(1):012011, 2020.

\bibitem{Bugg:2013ica}
W.~Bugg, Yu. Efremenko, and S.~Vasilyev.
\newblock {Large Plastic Scintillator Panels with WLS Fiber Readout;
  Optimization of Components}.
\newblock {\em Nucl. Instrum. Meth. A}, 758:91--96, 2014.

\bibitem{Jia-Ning2018Position-sensitive}
Jia-Ning Dong, Yun-Long Zhang, Zhi-Yong Zhang, Dong Liu, Zi-Zong Xu, Xiao-Lian
  Wang, and Shu-Bin Liu.
\newblock Position-sensitive plastic scintillator detector with wls-fiber
  readout.
\newblock {\em Nuclear Science and Techniques}, 29(8), 2018.

\bibitem{Adam:2007ex}
T.~Adam et~al.
\newblock {The OPERA experiment target tracker}.
\newblock {\em Nucl. Instrum. Meth. A}, 577:523--539, 2007.

\bibitem{MINOS:2002xlc}
P.~Adamson et~al.
\newblock {The MINOS scintillator calorimeter system}.
\newblock {\em IEEE Trans. Nucl. Sci.}, 49:861--863, 2002.

\bibitem{Wang:2021ejh}
Ya-Ping Wang, Chao Hou, Xiang-Dong Sheng, Shao-Hui Feng, Hong-Kui Lv, Jia Liu,
  Jing Zhao, Xiao-Peng Zhang, and Quan-Bu Gou.
\newblock {Testing and analysis of the plastic scintillator units for
  LHAASO-ED}.
\newblock {\em Rad. Det. Tech. Meth.}, 5(4):513--519, 2021.

\bibitem{LHAASO:2021awk}
F.~Aharonian et~al.
\newblock {Performance test of the electromagnetic particle detectors for the
  LHAASO experiment}.
\newblock {\em Nucl. Instrum. Meth. A}, 1001:165193, 2021.

\bibitem{Evans:2013pka}
Justin Evans.
\newblock {The MINOS Experiment: Results and Prospects}.
\newblock {\em Adv. High Energy Phys.}, 2013:182537, 2013.

\bibitem{Orsi:2007zz}
Silvio Orsi.
\newblock {PAMELA: A payload for antimatter matter exploration and light nuclei
  astrophysics}.
\newblock {\em Nucl. Instrum. Meth. A}, 580:880--883, 2007.

\bibitem{Andreev:2004uy}
V.~Andreev et~al.
\newblock {A high granularity scintillator hadronic-calorimeter with SiPM
  readout for a linear collider detector}.
\newblock {\em Nucl. Instrum. Meth. A}, 540:368--380, 2005.

\bibitem{Thompson:2022ufx}
David~J. Thompson and Colleen~A. Wilson-Hodge.
\newblock {Fermi Gamma-ray Space Telescope}.
\newblock 10 2022.

\bibitem{PROCUREUR2018169}
S.~Procureur.
\newblock Muon imaging: Principles, technologies and applications.
\newblock {\em Nuclear Instruments and Methods in Physics Research Section A:
  Accelerators, Spectrometers, Detectors and Associated Equipment},
  878:169--179, 2018.
\newblock Radiation Imaging Techniques and Applications.

\bibitem{Morishima:2017ghw}
Kunihiro Morishima et~al.
\newblock {Discovery of a big void in Khufu's Pyramid by observation of
  cosmic-ray muons}.
\newblock {\em Nature}, 552(7685):386--390, 2017.

\bibitem{Zenoni:2014kva}
Aldo Zenoni.
\newblock {Historical building stability monitoring by means of a cosmic ray
  tracking system}.
\newblock In {\em {4th International Conference on Advancements in Nuclear
  Instrumentation Measurement Methods and their Applications}}, IEEE
  Nucl.Sci.Symp.Conf.Rec., 3 2014.

\bibitem{Marteau:2012zv}
J.~Marteau, D.~Gibert, N.~Lesparre, F.~Nicollin, P.~Noli, and F.~Giacoppo.
\newblock {Muons tomography applied to geosciences and volcanology}.
\newblock {\em Nucl. Instrum. Meth. A}, 695:23--28, 2012.

\bibitem{Oguri:2014gta}
S.~Oguri, Y.~Kuroda, Y.~Kato, R.~Nakata, Y.~Inoue, C.~Ito, and M.~Minowa.
\newblock {Reactor antineutrino monitoring with a plastic scintillator array as
  a new safeguards method}.
\newblock {\em Nucl. Instrum. Meth. A}, 757:33--39, 2014.

\bibitem{Georgadze:2016ufb}
A.~Sh. Georgadze, V.~M. Pavlovych, O.~A. Ponkratenko, and D.~A. Litvinov.
\newblock {A remote reactor monitoring with plastic scintillation detector}.
\newblock 10 2016.

\bibitem{Scovell:2013hva}
P.~R. Scovell et~al.
\newblock {Low background anti-neutrino monitoring with an innovative composite
  solid scintillator detector}.
\newblock In {\em {2013 IEEE Nuclear Science Symposium and Medical Imaging
  Conference and Workshop on Room-Temperature Semiconductor Detectors}}, 2013.

\bibitem{Capozzi:2020cxm}
Francesco Capozzi, Eligio Lisi, and Antonio Marrone.
\newblock {Mapping reactor neutrino spectra from TAO to JUNO}.
\newblock {\em Phys. Rev. D}, 102(5):056001, 2020.

\bibitem{geant4-AGOSTINELLI2003250}
S.~Agostinelli et~al.
\newblock Geant4—a simulation toolkit.
\newblock {\em Nuclear Instruments and Methods in Physics Research Section A:
  Accelerators, Spectrometers, Detectors and Associated Equipment},
  506(3):250--303, 2003.

\bibitem{Riggi:2010zz}
S.~Riggi, P.~La~Rocca, E.~Leonora, D.~Lo~Presti, G.~S. Pappalardo, F.~Riggi,
  and G.~V. Russo.
\newblock {Geant4 simulation of plastic scintillator strips with embedded
  optical fibers for a prototype of tomographic system}.
\newblock {\em Nucl. Instrum. Meth. A}, 624:583--590, 2010.

\bibitem{Wenzhen2013Geant4}
Wenzhen XU, Yanfen LIU, Zongquan TAN, Ran XIAO, Wei KONG, and Bangjiao YE.
\newblock Geant4 simulation of plastic scintillators for a prototype usr
  spectrometer.
\newblock {\em Nuclear Science and Techniques}, 24(4), 2013.

\bibitem{Lecoq:2020itu}
P.~Lecoq.
\newblock {\em {Scintillation Detectors for Charged Particles and Photons}},
  pages 45--89.
\newblock Springer, Cham, 2020.

\bibitem{Min2023Performance}
Min Li, Zhi-Min Wang, Cai-Mei Liu, Pei-Zhi Lu, Guang Luo, Yuen-Keung Hor,
  Jin-Chang Liu, and Chang-Gen Yang.
\newblock Performance of compact plastic scintillator strips with wavelength
  shifting fibers using a photomultiplier tube or silicon photomultiplier
  readout.
\newblock {\em Nuclear Science and Techniques}, 34(2), 2023.

\bibitem{Yang:2022awv}
Hang Yang, Guang Luo, Tao Yu, Shihan Zhao, Biying Hu, Zhencheng Huang, Han
  Shen, Lili Yang, Yu~Chen, and Jian Tang.
\newblock {MuGrid: A scintillator detector towards cosmic muon absorption
  imaging}.
\newblock {\em Nucl. Instrum. Meth. A}, 1042:167402, 2022.

\bibitem{Haotang:2014}
Hoton technology co. beijing hoton nuclear technology co., ltd., 2014.

\bibitem{Tur:2009en}
Clarisse Tur, Vladimir Solovyev, and Jeremy Flamanc.
\newblock {Temperature characterization of scintillation detectors using
  solid-state photomultipliers for radiation monitoring applications}.
\newblock {\em Nucl. Instrum. Meth. A}, 620:351--358, 2010.

\bibitem{Dietz-Laursonn:2016tpy}
Erik Dietz-Laursonn.
\newblock {\em {Detailed Studies of Light Transport in Optical Components of
  Particle Detectors}}.
\newblock PhD thesis, Aachen, Tech. Hochsch., 2016.

\bibitem{Qian:2021jlv}
Xiang-Li Qian, Hui-Ying Sun, Cheng Liu, Xu~Wang, and Olivier Martineau-Huynh.
\newblock {Simulation study on performance optimization of a prototype
  scintillation detector for the GRANDProto35 experiment}.
\newblock {\em Nucl. Sci. Tech.}, 32(5):51, 2021.

\bibitem{Gaisser:2016ddr}
Thomas Gaisser.
\newblock {Cosmic-Ray Showers Reveal Muon Mystery}.
\newblock {\em APS Physics}, 9:125, 2016.

\bibitem{Gaisser:2016uoy}
Thomas~K. Gaisser, Ralph Engel, and Elisa Resconi.
\newblock {\em {Cosmic Rays and Particle Physics}: {2nd Edition}}.
\newblock Cambridge University Press, 6 2016.

\bibitem{Shukla:2016nio}
Prashant Shukla and Sundaresh Sankrith.
\newblock {Energy and angular distributions of atmospheric muons at the Earth}.
\newblock {\em Int. J. Mod. Phys. A}, 33(30):1850175, 2018.

\bibitem{citeonsemi}
https://www.onsemi.com/.

\end{thebibliography}
\end{document}